\documentclass[pr, 12pt, superscriptaddress]{revtex4}  
\usepackage{amssymb}
\usepackage{amsmath}
\usepackage{bbm}
\usepackage{graphicx}
\usepackage{color}

\begin{document}

\author{Philipp Werner}
\affiliation{Department of Physics, University of Fribourg, 1700 Fribourg, Switzerland}
\author{Markus Lysne}
\thanks{Currently at Sintef Energy Research}
\affiliation{Department of Physics, University of Fribourg, 1700 Fribourg, Switzerland}
\author{Yuta Murakami}
\affiliation{Center for Emergent Matter Science, RIKEN, Saitama 351-0198, Japan}

\title{High harmonic generation in altermagnets}

\date{\today}

\hyphenation{}

\begin{abstract} 
We study high harmonic generation in altermagnetic metals with and without spin-orbit coupling. The altermagnetism manifests itself in the magnetic field dependence of the low harmonics associated with intra-band dynamics. Spin-orbit coupling leads to additional higher energy peaks and plateau structures originating from inter-band transitions. While the pure altermagnet or spin-orbit system exhibits no circular dichroism in the high-harmonic response, an altermagnetic system with spin-orbit coupling shows such a dichroism. We also analyze the  spin currents and their high harmonic spectrum.  
\end{abstract}

\maketitle


\section{Introduction}

High-harmonic generation (HHG) is a nonlinear optical process, where a system driven by a laser field with frequency $\Omega$ emits radiation at (potentially very high) multiples of this fundamental frequency. HHG has been extensively studied in atomic gases \cite{Ferray1988}, where the phenomenon can be understood in terms of the three step model describing tunnel ionization, acceleration in free space, and recombination \cite{Corkum1993,Lewenstein1994}. More recently, HHG has been realized in liquids \cite{Luu2018}, semiconductors \cite{Ghimire2011,Schubert2014,Luu2015,Vampa2015,Langer2016,Yoshikawa2017,Liu2017,Kaneshima2018,Yoshikawa2019}, and even strongly correlated materials \cite{Bionta2021,Uchida2022,Alcala2022}. Also in the solid state context, an adaptation of the three step model helps to understand the HHG process \cite{Vampa2015prb,Murakami2018,Imai2020,Murakami2021,Li2023}. Here, the tunnel ionization is replaced by electron-holon production and the evolution of the charge carriers takes place within a bandstructure. Because the properties of the bandstructure are reflected in the HHG spectrum, several studies suggested to use HHG to map out band dispersions or to detect the topological nature of materials \cite{Vampa2015prl,Luu2015,Luu2018top,Li2020}. 

In recent years, altermagnets have emerged as a new class of collinear magnets with zero net magnetization but a spin-split bandstructure and $d$-wave-like spin-split Fermi surfaces \cite{Naka2019,Smejkal2022a,Smejkal2022}. The spin splitting leads to spin polarized currents or even pure spin currents, and various interesting nonlinear effects on transport have been discussed \cite{Farajollahpour2024}. Here, we consider a simple model of an altermagnetic metal to investigate effects of the altermagnetic bandstructure on HHG. We show that the magnetic field dependence of the HHG spectrum allows to reveal the $d$-wave nature of the Fermi surfaces, while in systems with spin-orbit coupling, a circular dichroism in the high-harmonic response can be an indication for altermagnetism. 

The paper is organized as follows. Section~\ref{sec:model} introduces the model. Section~\ref{sec:results_alpha0} analyzes HHG from the charge current and the high-harmonic components of the spin currents in a pure altermagnet model, while Sec.~\ref{sec:results_alpha1} shows results for a model with altermagnetic spin splitting and spin-orbit coupling. Our conclusions are presented in Sec.~\ref{sec:conclusions}.

\section{Model}
\label{sec:model}

We consider a simple model of a two-dimensional altermagnetic metal 
$H(t)=\sum_{\boldsymbol{k}} \Psi^\dagger_{\boldsymbol{k}} h_{\boldsymbol{k}}(t) \Psi_{\boldsymbol{k}}$, with $\Psi_{\boldsymbol{k}}^\dagger=(c_{{\boldsymbol{k}}\uparrow}^\dagger,c_{{\boldsymbol{k}},\downarrow}^\dagger)$ a spinor for the creation operators at momentum ${\boldsymbol{k}}$, and $h_{\boldsymbol{k}}=h^\text{alter}_{\boldsymbol{k}}+h^\text{SOC}_{\boldsymbol{k}}$ composed of an altermagnet term $h^\text{alter}_{\boldsymbol{k}}$ with anisotropic spin-dependent hoppings \cite{Lee2024} (plus a Zeeman field), and $h^\text{SOC}_{\boldsymbol{k}}$ representing a Rashba spin-orbit coupling \cite{Lysne2020}. The explicit expressions for the two terms are
\begin{align}
&h^\text{alter}_{\boldsymbol{k}}(t)=(t_0[2-\cos(k_x(t))-\cos(k_y(t))]-\mu)\sigma_0 
+2t_a[\cos(k_x(t))-\cos(k_y(t))]\sigma_3+b\sigma_3, \label{eq_alter}\\
&h^\text{SOC}_{\boldsymbol{k}}(t)=-\alpha \sin(k_y(t)) \sigma_1 +\alpha \sin(k_x(t)) \sigma_2, \label{eq_soc}
\end{align}
where the $\sigma_{0,1,2,3}$ denote the identity and Pauli matrices in spin space. $t_0$ controls the spin-averaged bandwidth, $t_a$ the spin splitting, $\mu$ is the chemical potential, $b$ the Zeeman field, and $\alpha$ the spin-orbit coupling. In the presence of an electric field, the momenta $k_{x,y}$ can become time-dependent (Peierls substitution \cite{Peierls1933}), as discussed below. The unit of energy is set to $t_0=1$, which means that the unit of time is $\hbar/t_0$ ($\hbar=1$ in the following). 

To simulate the current and HHG spectrum produced by an electric field pulse, we use a gauge with a pure vector potential, so that the electric field ${\bf E}(t)$ is determined by the time derivative of the vector potential ${\bf A}(t)$: ${\bf E}(t)=-\partial_t {\bf A}(t)$. We furthermore assume that the wave length of the field is much larger than the lattice spacing (dipole approximation). In this case, the effect of the electric field can be incorporated into $h_{\boldsymbol{k}}$ by substituting ${\boldsymbol{k}}\rightarrow {\boldsymbol{k}}+{\bf A}(t)$, where we have set the charge $q$ of the electron to $q=-1$ and the lattice spacing to $a=1$. The form of the pulse is 
\begin{equation}
E_i(t)=E_{0,i} \sin(\Omega(t-t_\text{avg}+\phi_i))\cos^8(\Omega(t-t_\text{avg})/2M), \quad (0<t<MT) \label{eq_field}
\end{equation}
with $E_{0,i}$ the strength of the field component $i$, $\phi_i$ the corresponding phase (which is zero for linear polarization) and $\Omega$ the pulse frequency. The pulse contains $M$ cycles of period $T$, and it is centered at $t_\text{avg}=MT/2$. For $t>MT$, the field is zero. In calculations with $b\ne 0$, we assume that the Zeeman term dominates the effect of the magnetic field on the current and high-harmonic spectrum. In practice this can be guaranteed by choosing $b$ such that the cyclotron frequency $\omega_c$ ($=b$) is small compared to $\Omega$. 

The time evolution of the system's momentum-dependent density matrix, $\rho(\boldsymbol{k},t)$,  is described by the Von-Neumann equation
$$
	\frac{d}{dt}\rho(\boldsymbol{k}, t) = -i[ h_{\boldsymbol{k}}(t), \rho(\boldsymbol{k}, t) ],
$$
which is solved by a commutator-free expansion, as detailed in Ref.~\cite{Alvermann2011}. The high-harmonic spectrum for light polarized along $i$ is calculated from the Fourier transform of the time derivative of the charge current 
\begin{equation}
J_i(t)=-\frac{1}{N_{\boldsymbol{k}}}\sum_{\boldsymbol{k}}\text{Tr}[\rho(\boldsymbol{k},t)v_i(\boldsymbol{k},t)], \quad v_i(\boldsymbol{k},t)=\partial h_{\boldsymbol{k}}(t)/\partial {\boldsymbol{k}_i}, \label{eq_chargecurrent}
\end{equation}
as $|\omega J_{i}(\omega)|^2$ (neglecting the potential contribution from an induced time-dependent magnetization) \cite{Schueler2021}. Similarly, we define the high-harmonic spectrum of the spin current as $|\omega J_{ij}(\omega)|^2$, with
\begin{equation}
J_{ij}(t)=-\frac{1}{N_{\boldsymbol{k}}}\sum_{\boldsymbol{k}}\text{Tr}[\rho(\boldsymbol{k},t)v_{ij}(\boldsymbol{k},t)], \quad v_{ij}(\boldsymbol{k},t)=\frac{1}{2}[\sigma_i v_j(\boldsymbol{k},t)+v_{j}(\boldsymbol{k},t)\sigma_i]. \label{eq_spincurrent}
\end{equation}

\section{Results}

\subsection{Altermagnet ($\alpha=0$)}
\label{sec:results_alpha0}

\begin{figure}[t]
\begin{center}
\includegraphics[angle=-90, width=0.30\columnwidth]{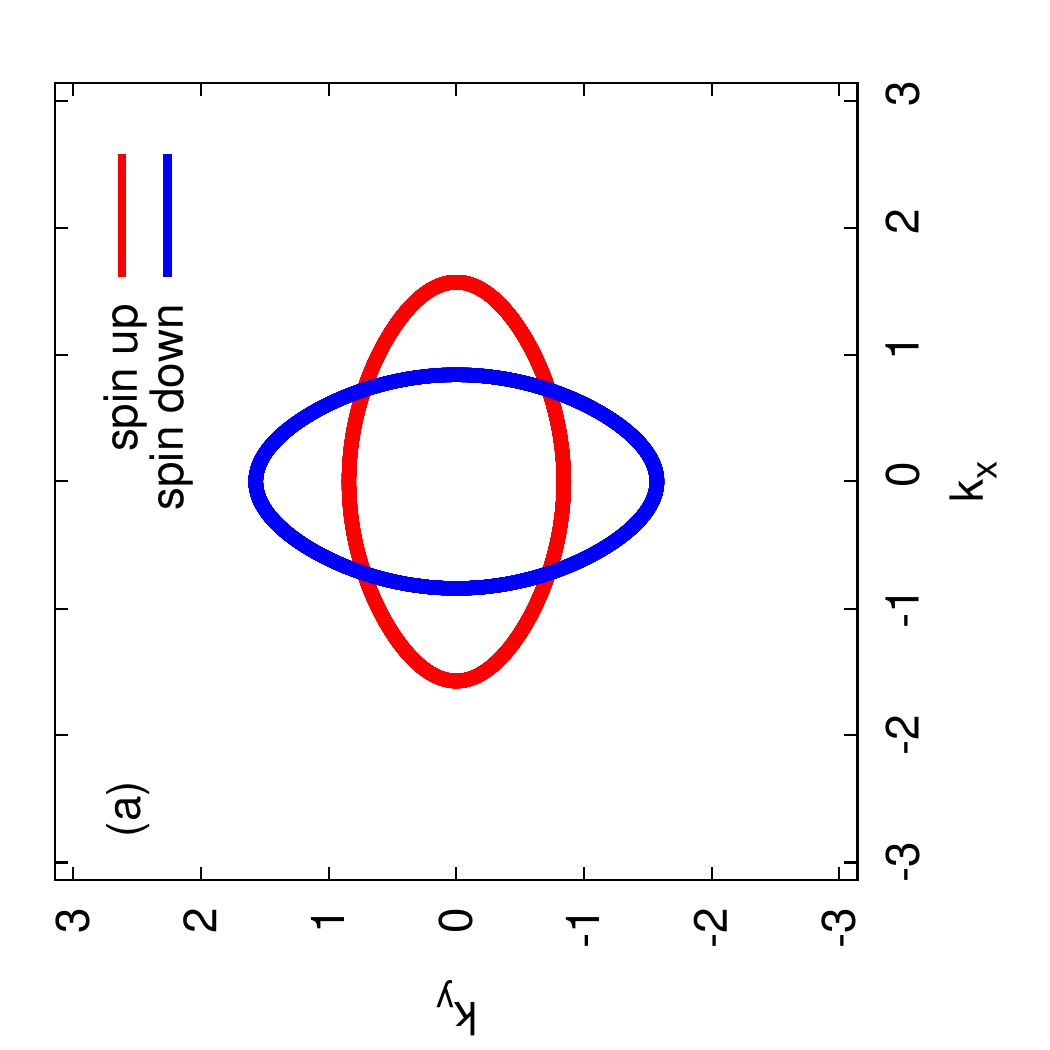}\hfill
\includegraphics[angle=-90, width=0.30\columnwidth]{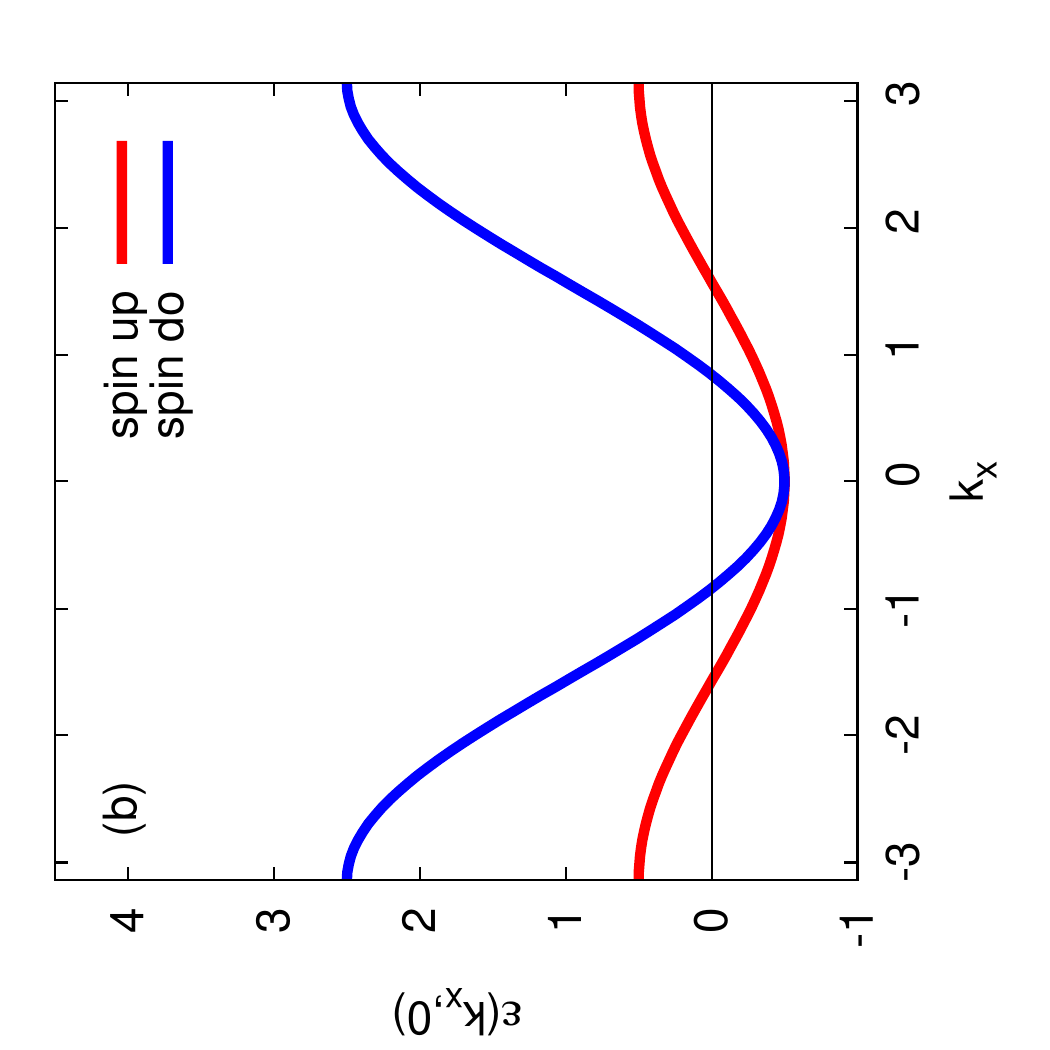}\hfill
\includegraphics[angle=-90, width=0.30\columnwidth]{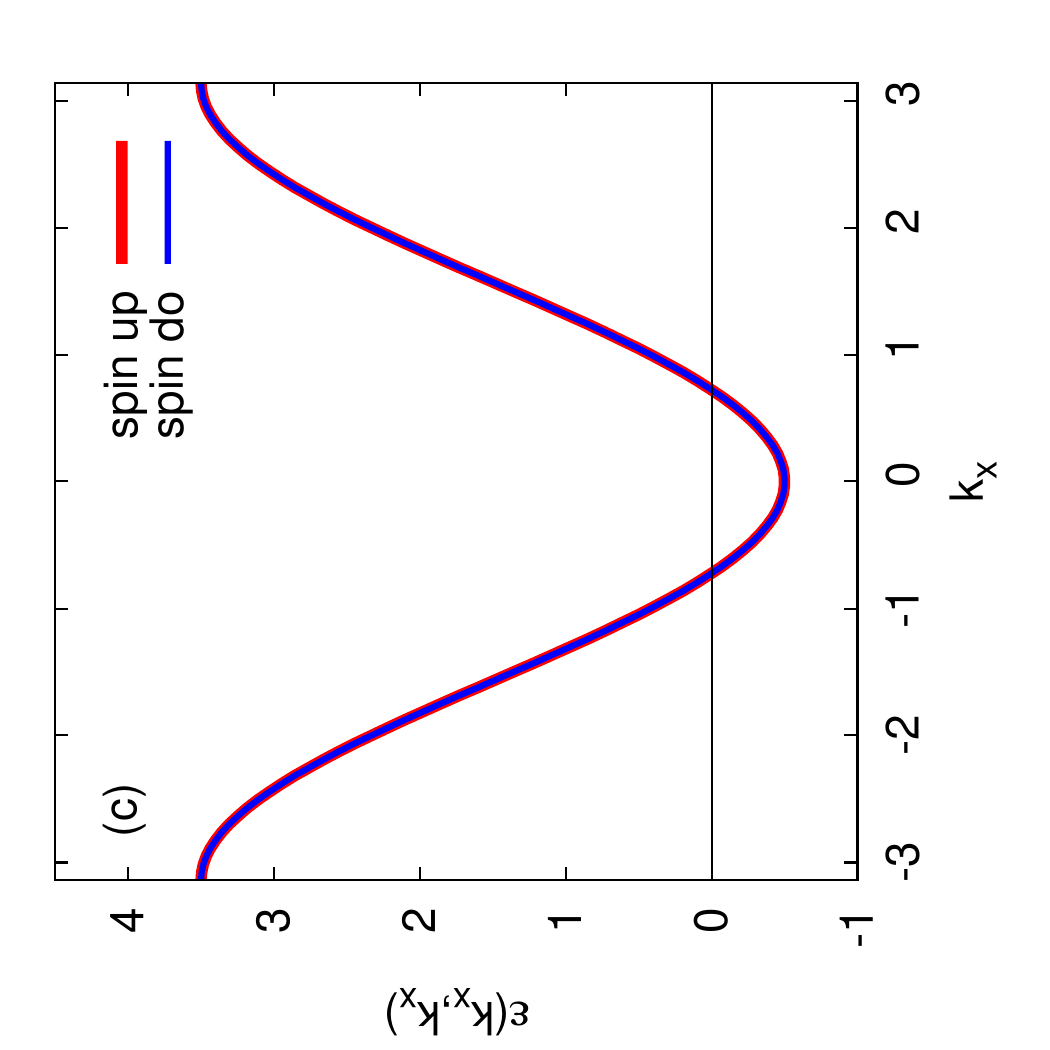}
\caption{Fermi surfaces (a) and spin-up/down bands for $k_y=0$ (b) and $k_x=k_y$ (c). The model parameters are $t_a=0.25$, $b=0$, and $\mu=0.5$. The states with $\varepsilon({\boldsymbol{k}})<0$ are initially occupied.
}
\label{fig_FS_bands}
\end{center}
\end{figure}

We start by considering the pure altermagnet model, $h_{\boldsymbol{k}}=h_{\boldsymbol{k}}^\text{alter}$, setting $t_0=1$, $t_a=0.25$ and $\mu=0.5$. 
For these parameters, the spin-up Fermi surface is elongated along $k_x$ and the spin-down Fermi surface elongated along $k_y$, as shown in Fig.~\ref{fig_FS_bands}(a). Panel (b) plots the dispersion $\varepsilon({\boldsymbol{k}})$ for $k_y=0$, where the spin-down band is three times wider than the spin-up band. Along the diagonals $k_x=k_y$, the spin-up and -down dispersions are degenerate (panel (c)). 

The HHG spectra are calculated for 20-cycle pulses ($M=20$) with frequency $\Omega=0.1$ and linear polarization, unless otherwise stated. We use a $200\times 200$ momentum grid  in the calculations without spin-orbit coupling, i.e. $N_{\boldsymbol{k}}=200^2$ in Eqs.~\eqref{eq_chargecurrent} and \eqref{eq_spincurrent}. The initial temperature is set to $T=1/30$.

\subsubsection{Charge current high-harmonic spectrum}

\begin{figure}[t]
\begin{center}
\includegraphics[angle=0, width=0.32\columnwidth]{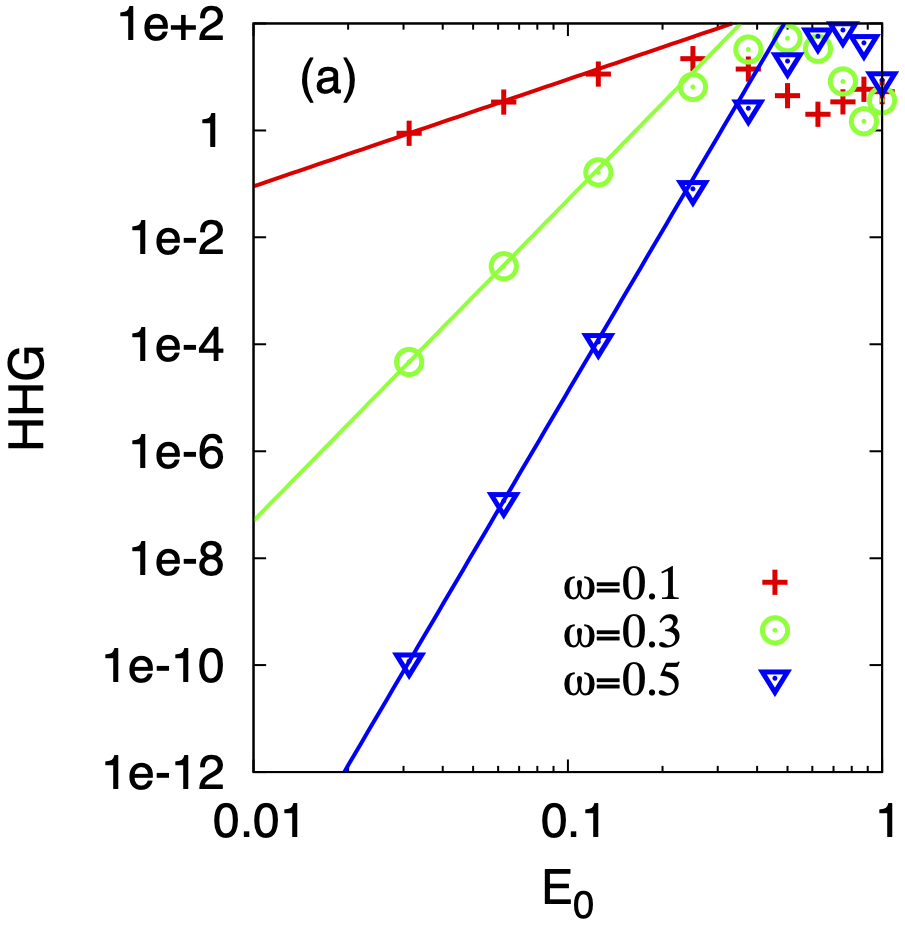}\hspace{15mm}
\includegraphics[angle=0, width=0.37\columnwidth]{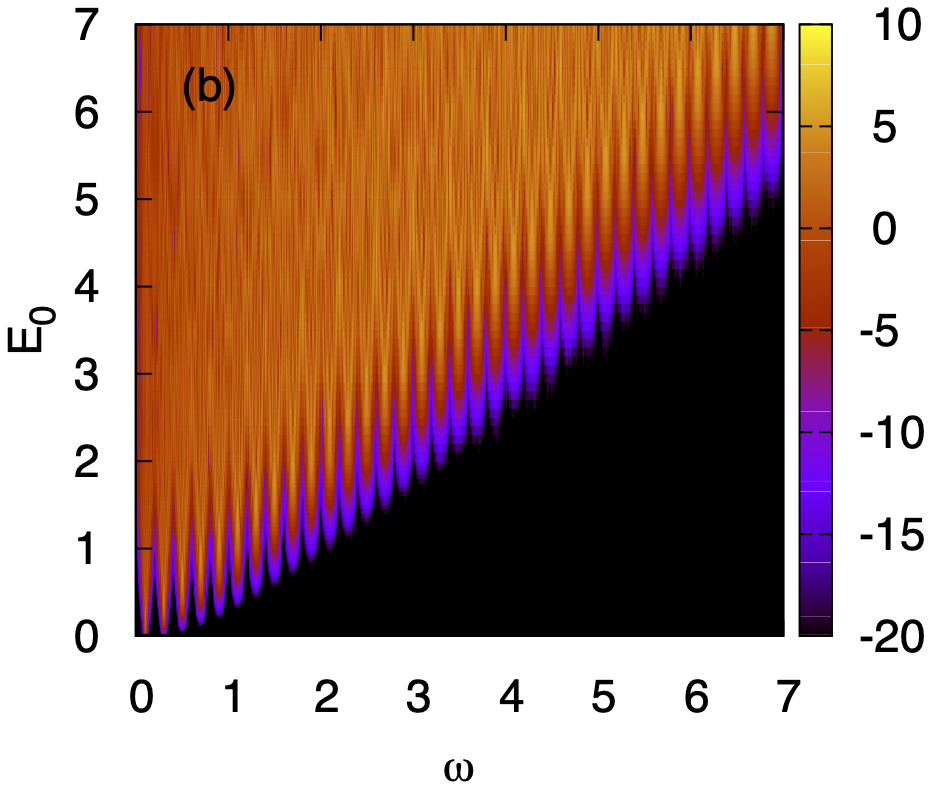}
\caption{Panel (a): Scaling of the HHG intensity for weak fields with polarization along $x$. The red, green, and blue symbols show the amplitude of the spectrum for the first, third and fifth harmonic ($\omega=n\Omega$, $n=1,3,5$, $\Omega=0.1$), while the lines are proportional to $E_0^2$, $E_0^6$ and $E_0^{10}$. Panel (b): Logarithm of the HHG spectrum as a function of $E_0$, for polarization along $x$.}
\label{fig_hhg}
\end{center}
\end{figure}

\begin{figure}[t]
\begin{center}
\includegraphics[angle=0, height=0.22\columnwidth]{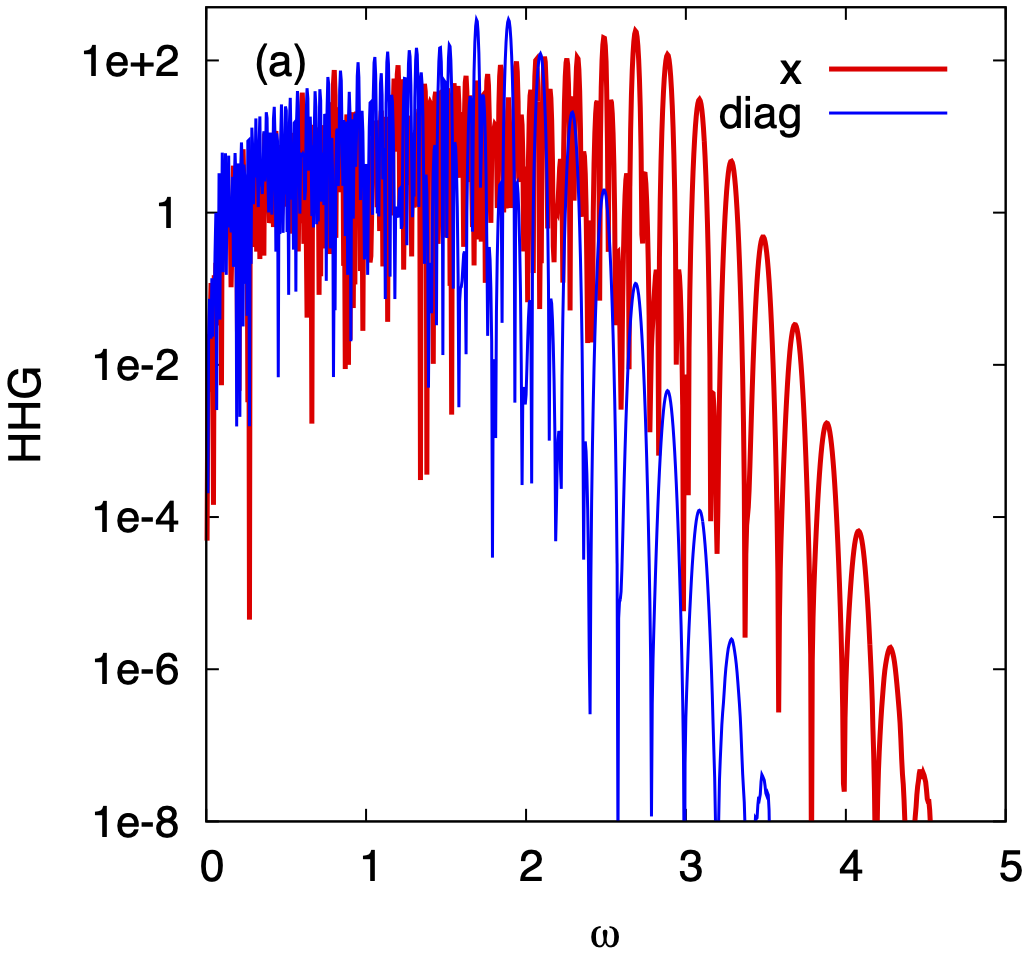}\hfill
\includegraphics[angle=0, height=0.22\columnwidth]{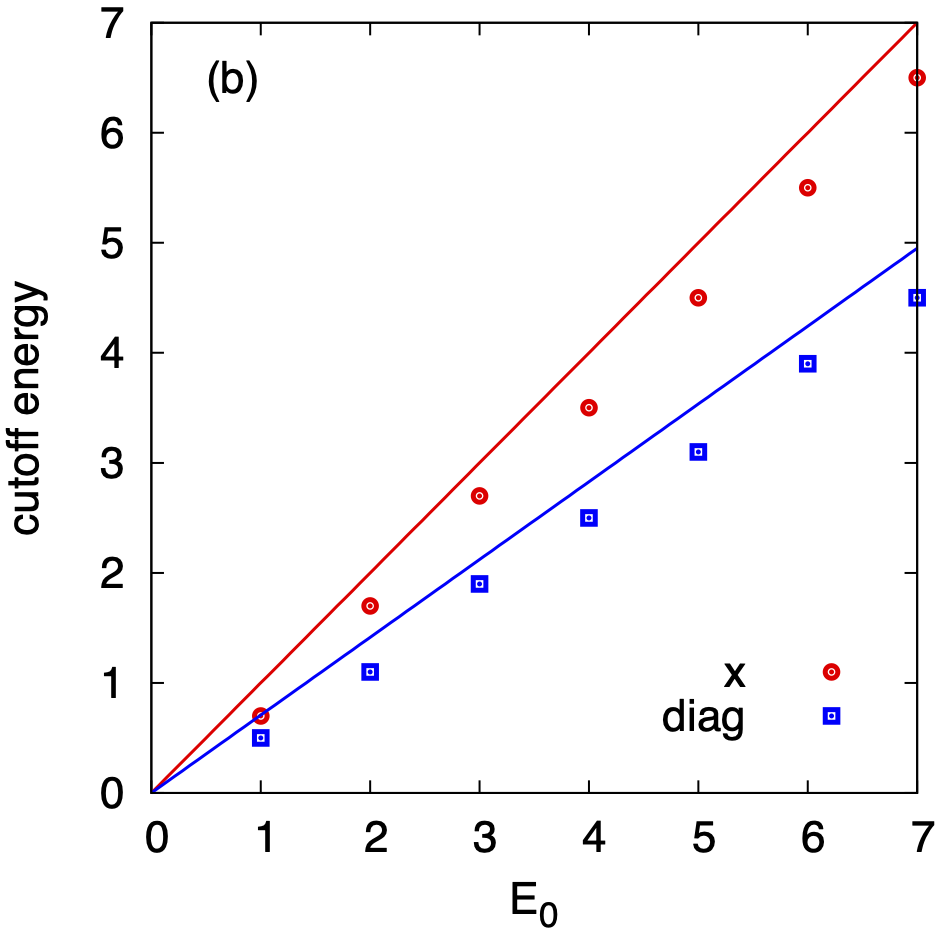} \hfill
\includegraphics[angle=0, height=0.22\columnwidth]{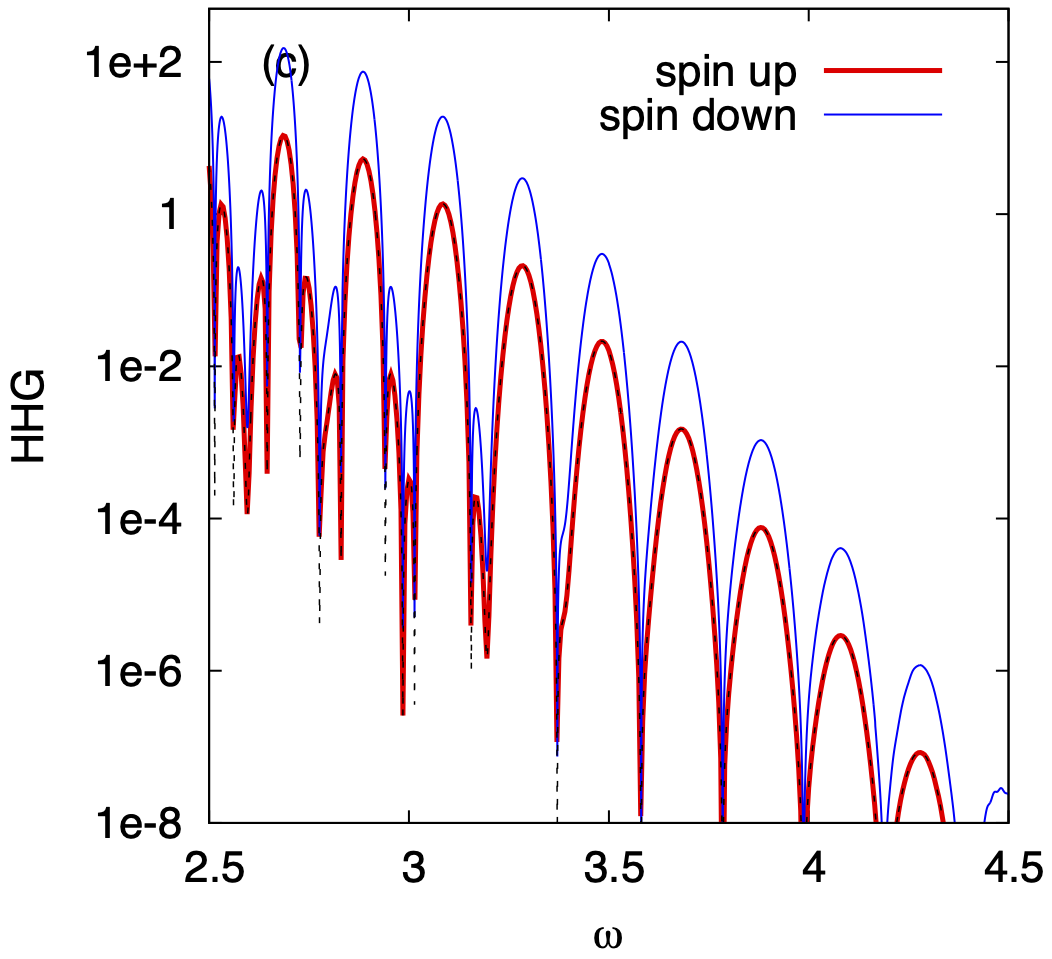}\hfill
\includegraphics[angle=0, height=0.22\columnwidth]{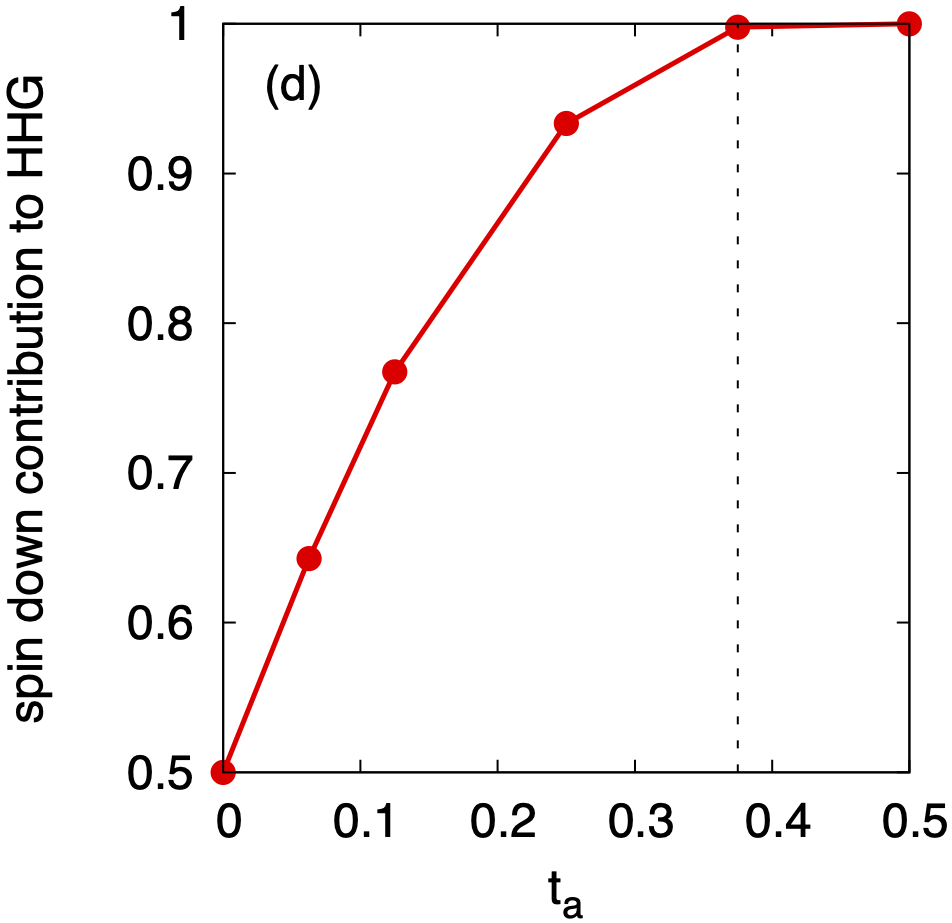}
\caption{(a) HHG spectrum for $E_0=3$ and pump polarization along $x$ (red) and along the diagonal (blue). (b) cutoff energy as a function of $E_0$ for pump polarization along $x$ (red) and along the diagonal (blue). (c) HHG spectrum for the system with only a spin-up/down band, for $E_0=3$ and pump polarization along $x$. The black dashed line shows the spin-down spectrum divided by a factor 14. Panels (a-c) show results for $t_a=0.25$.    
(d) Relative contribution of the spin-down current to the total HHG signal, as a function of $t_a$. The dashed line at $t_a=0.375$ indicates the change from closed to open Fermi surfaces.  
}
\label{fig_cutoff}
\end{center}
\end{figure}

If $\alpha=0$, spin is conserved, and the features of the HHG spectrum originate from intra-band currents. 
For small fields, the amplitudes of the low harmonics $\omega=n\Omega$ ($n=1,3,5, \ldots$) grow proportional to $E_0^{2n}$, as shown in Fig.~\ref{fig_hhg}(a) for field polarization along $x$. This is consistent with the expected scaling of the nonlinear current contributions $j^{(n)}\propto E_0^n$. 
For strong fields polarized along $x$, the HHG spectrum has a plateau up to a cutoff energy $\omega_\text{cut}\approx E_0$ \cite{Kemper2013}, see Fig.~\ref{fig_hhg}(b). Because of the absence of a damping term, the spectrum in the plateau region has a complicated structure, with some constructive and destructive interference effects at even multiples of $\Omega$, but overall no well-defined harmonics. In the vicinity of $\omega_\text{cut}$ and in the exponentially decaying region, however, clear harmonics at odd multiples of $\Omega$ are observed (see Fig.~\ref{fig_cutoff}(a)), as expected from inversion symmetry \cite{Neufeld2019}. 
In strong electric fields, Wannier-Stark localization \cite{Kruchinin2018} leads to a series of states which are almost localized at the different sites, with energy spacing $E_0$ (the bond length $a$ is set to unity). In a real-space picture, the maximum energy gain from nearest-neighbor hopping in the field direction is $E_0$, which explains the cutoff $\omega_\text{cut}\approx E_0$ in the strong-field regime \cite{Murakami2018,Higuchi2014,Lysne2020a}.

Consistent with this quasi-local picture is the observation that $\omega_\text{cut}$ for fields polarized along the diagonal are smaller than for fields along $x$ or $y$ (Fig.~\ref{fig_cutoff}(a)), even though the dispersion of the bands is wider (Fig.~\ref{fig_hhg}(b)). Since the projection of the field with diagonal polarization onto the $x$ and $y$ axes is a factor $1/\sqrt{2}$ smaller, the energy gain from a single nearest-neighbor hopping process is correspondingly reduced. Indeed, as shown in Fig.~\ref{fig_cutoff}(b), the cutoff energy for pump polarization along the diagonal is approximately $1/\sqrt{2}$ times smaller than along $x$ (or $y$). 
We note, however, that for small field amplitudes ($E_0\lesssim 1$), this quasi-local picture is no longer appropriate, and longer trajectories contribute to the current.   

For $E_0 \gtrsim 1$, the same conclusions also hold for the model with $t_a=0$, i.e. without spin splitting. In fact, the different hopping amplitudes along $x$ and $y$ in the altermagnet have little effect on $\omega_\text{cut}$, they mainly affect the relative contributions of the two spin channels to the HHG signal. If we separately consider the HHG signal from the spin-up and spin-down current, then for field polarization along $x$, the spectrum produced by the spin-down electrons is a factor 14 times larger (for $t_a=0.25$) than the spectrum produced by the spin-up electrons (Fig.~\ref{fig_cutoff}(c)). This result for the strong-field regime is qualitatively consistent with Eq.~(\ref{eq_alter}) and Fig.~\ref{fig_FS_bands}, which imply that the down-spin hopping is three time larger along $x$ than the up-spin hopping. The ratio between the up- and down-spin contributions depends on the parameter $t_a$, which controls the shapes of the spin-dependent Fermi surfaces. For $t_a\rightarrow 0$, the two contributions become identical, while at $t_a = 0.375$, the topology of the Fermi surface changes from closed to open (1D-like). Beyond this value, the spin-up states along more and more $k_x$ cuts get fully occupied, so that the down-spin contribution completely dominates the current in the $x$ direction, as shown in Fig.~\ref{fig_cutoff}(d).

\begin{figure}[t]
\begin{center}
\includegraphics[angle=0, width=0.325\columnwidth]{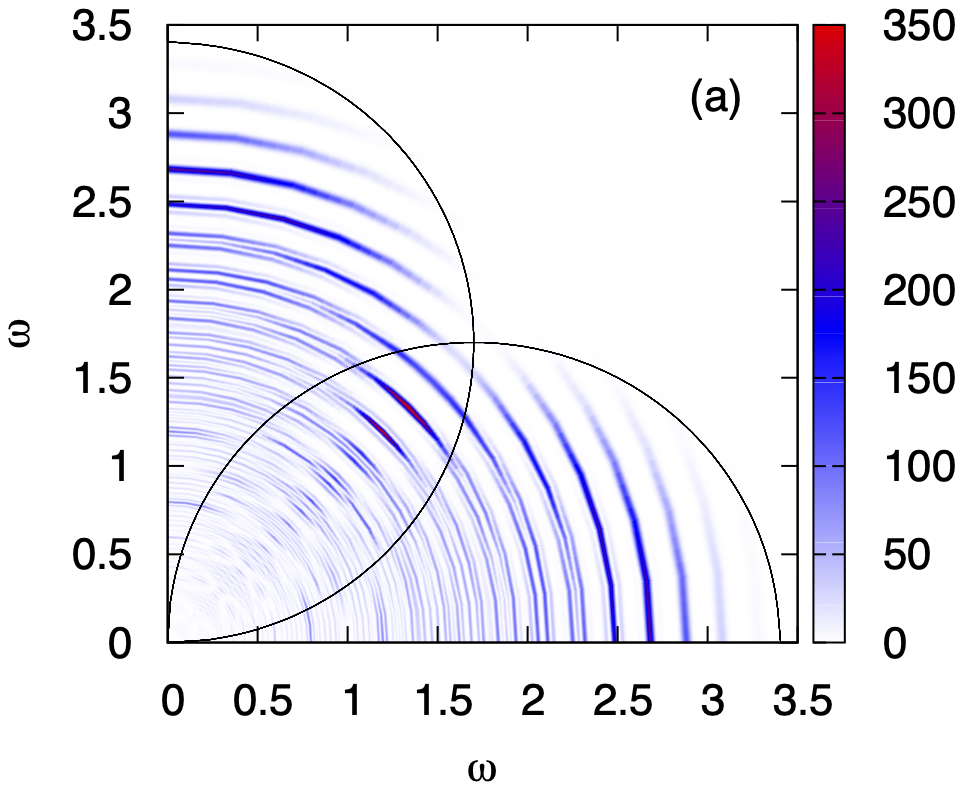}\hspace{3mm}
\includegraphics[angle=0, width=0.3\columnwidth]{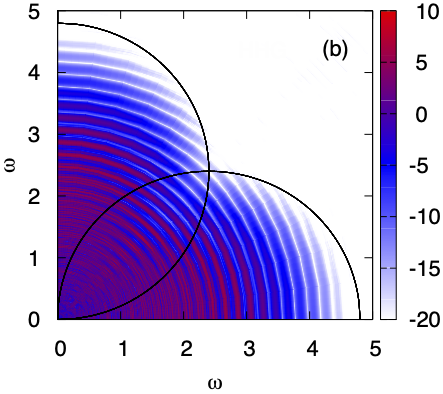}\hspace{3mm}
\includegraphics[angle=0, width=0.3\columnwidth]{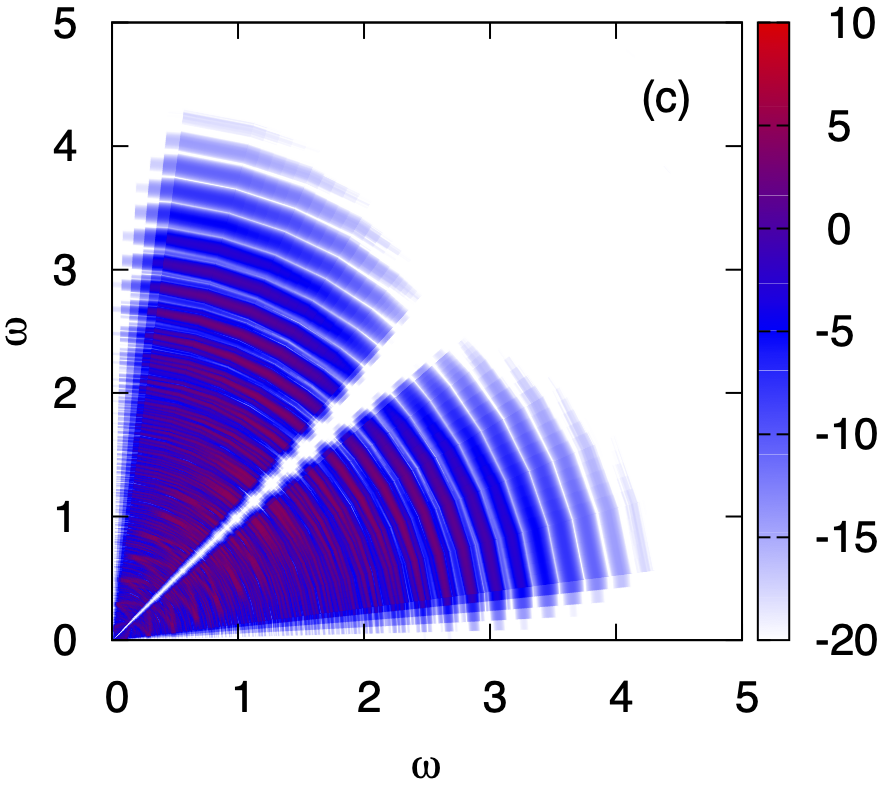}
\caption{HHG signal as a function of polarization angle for $E_0=3$. Panel (a) shows the spectra from the current component parallel to the field on a linear scale and panel (b) the logarithm of the spectra. The semi-circles with arbitrarily rescaled radii show the projection of the electric field onto the $x$ and $y$ axes. 
Panel (c) shows the logarithm of the spectra obtained from the current perpendicular to the applied field (in the azimuthal direction). 
}
\label{fig_hhg_polar_linear}
\end{center}
\end{figure}

The HHG spectra for $E_0=3$ are plotted as a function of polarization angle in Fig.~\ref{fig_hhg_polar_linear}. In panels (a) and (b), we show the HHG signal for light polarized parallel to the field (calculated from the current contribution parallel to the applied field). For generic field directions, the nonlinear response can lead to transverse currents, which in principle also contribute to the emitted power. 
The angle-dependent HHG spectra for the transverse current are shown in panel (c).

Focusing on the longitudinal current, we see from panels (a) and (b) that $\omega_\text{cut}$ is largest for $x$ or $y$ polarization, while the peak intensity is highest along the diagonal. The cutoff values roughly follow two semicircles centered on the $x$ and $y$ axes, respectively, consistent with the strength of the field projected onto these axes. The strong signal for polarization along the diagonal is in turn consistent with the larger bandwidth in the diagonal direction. Again, these properties are not specific to the altermagnet, 
and qualitatively similar spectra are obtained for $t_a=0$. 
The amplitude of the HHG signal is however larger in the altermagnetic system, which has different contributions from spin-up and spin-down currents. Since the HHG spectrum is quadratic in the current, the enhanced contribution from the spin component with larger bandwidth dominates the suppressed contribution from the spin component with reduced bandwidth, resulting in an overall stronger signal.

\subsubsection{Effect of Zeeman field}

\begin{figure}[t]
\begin{center}
\includegraphics[angle=-90, width=0.32\columnwidth]{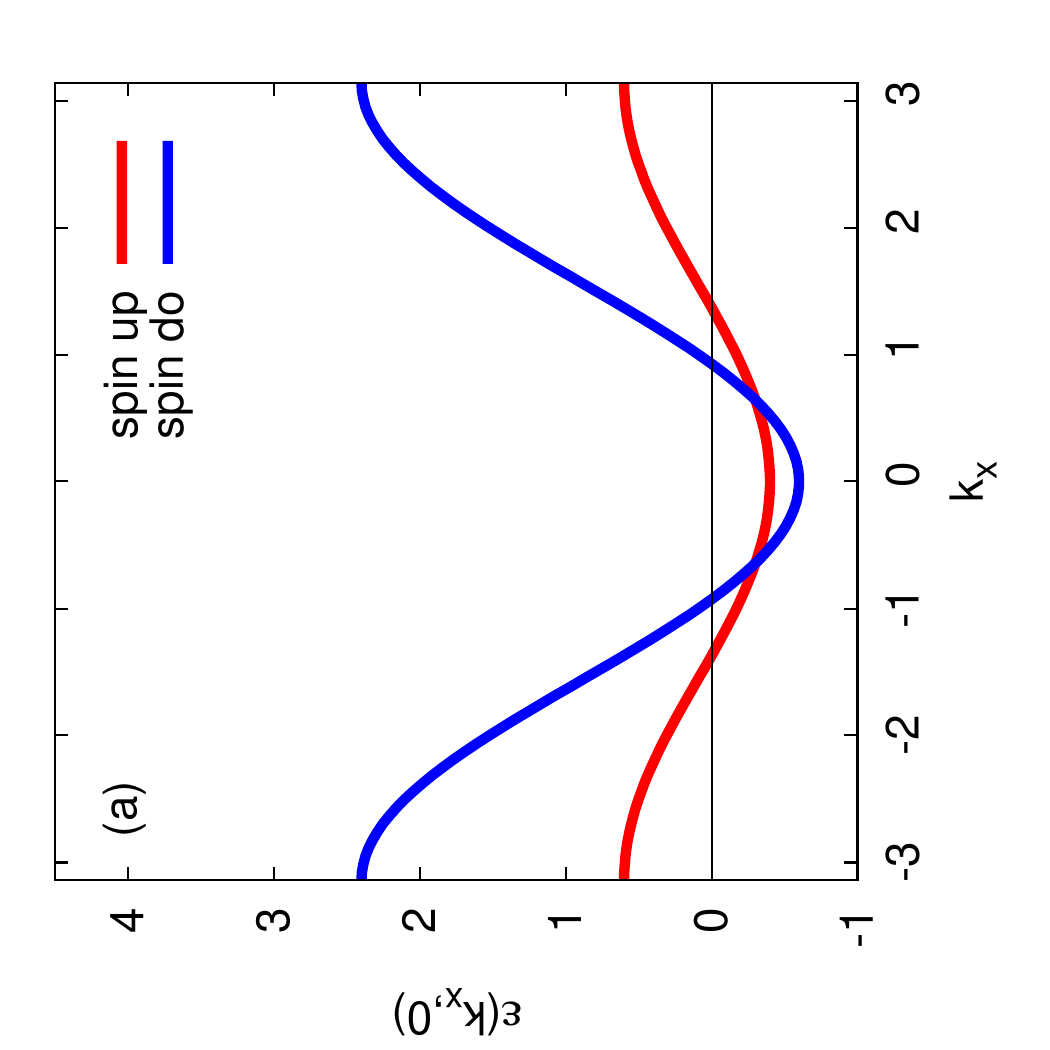}
\includegraphics[angle=-90, width=0.32\columnwidth]{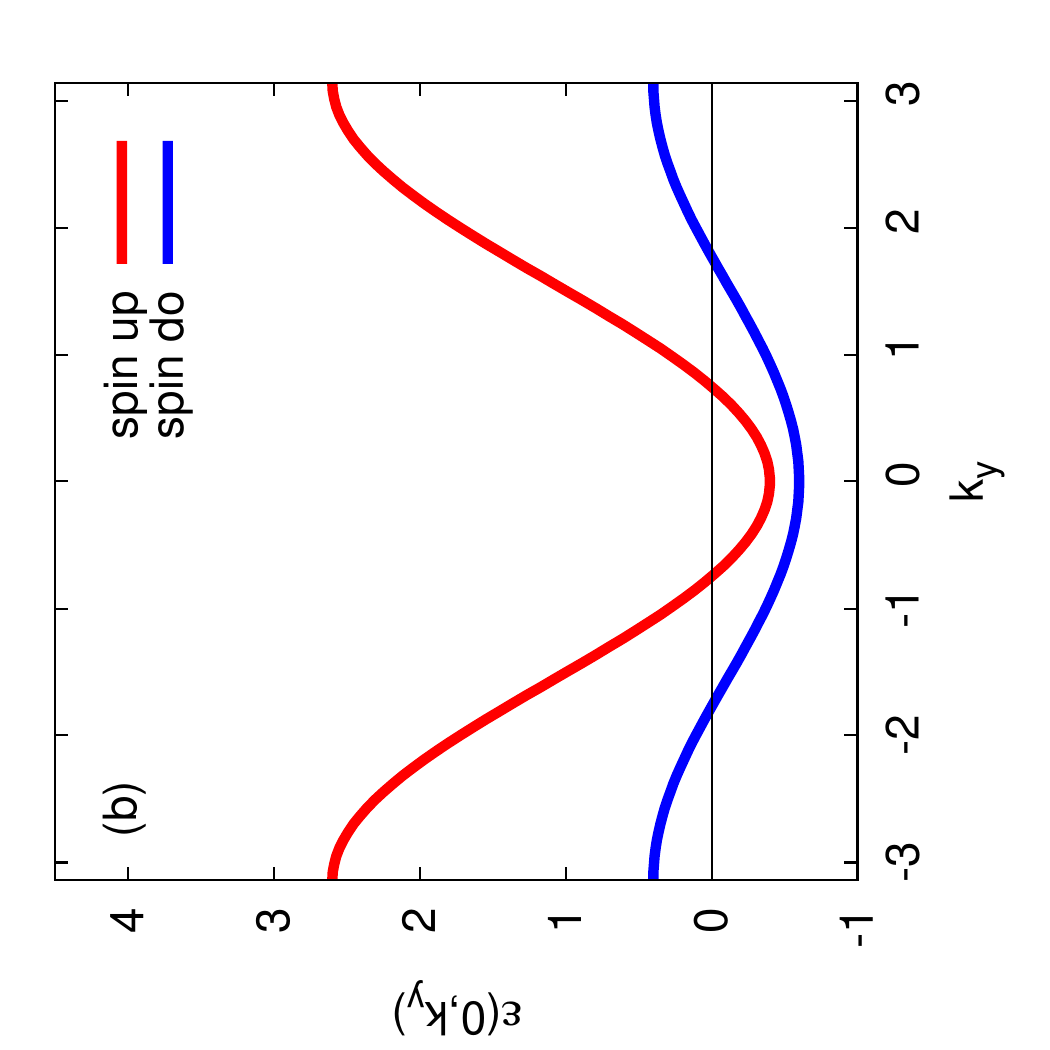}
\includegraphics[angle=-90, width=0.33\columnwidth]{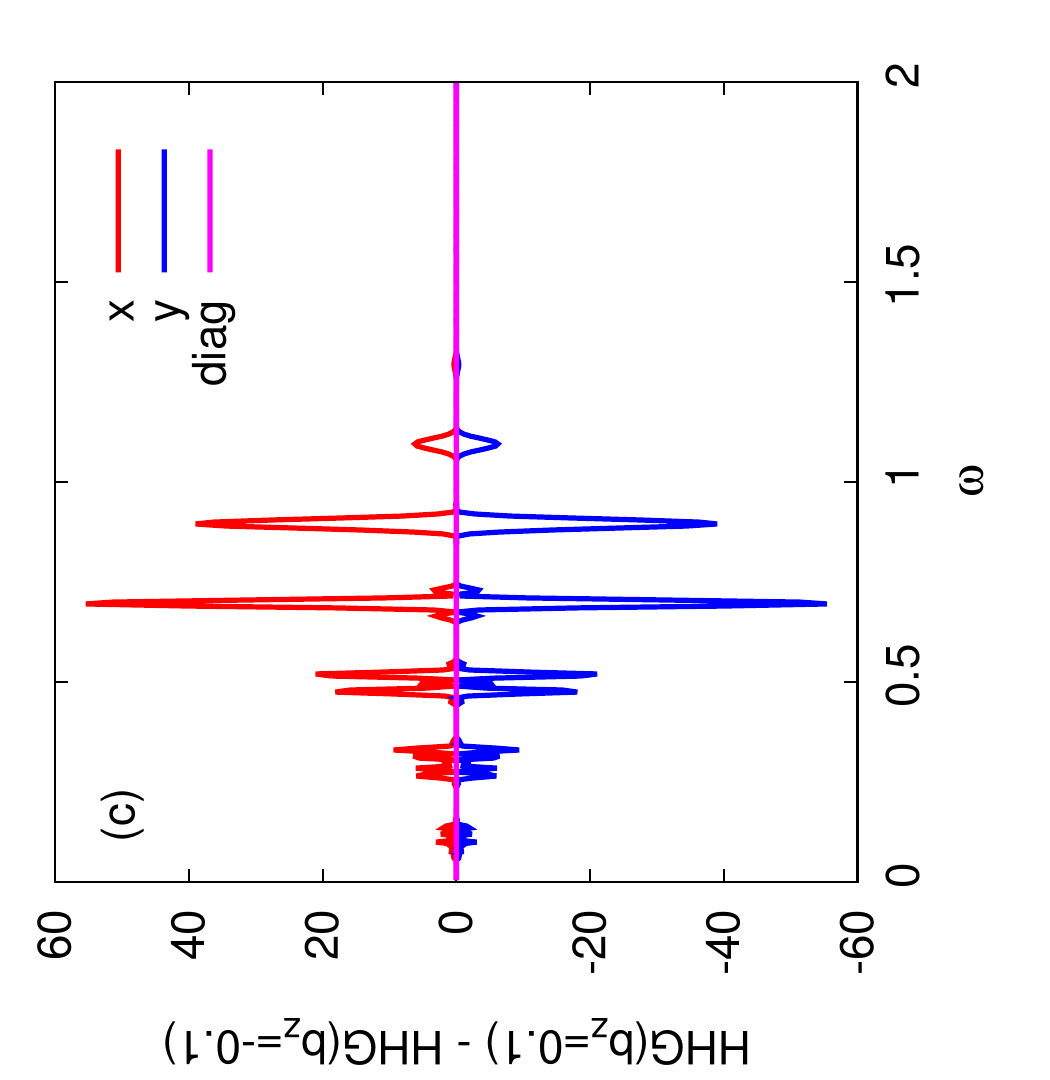}
\caption{Bands at $k_y=0$ (a) and $k_x=0$ (b) for $b=0.1$. Panel (c) shows the difference in the HHG spectrum for $b=\pm 0.1$, for a pump field with $E_0=3$ and polarization along $x$. The states with $\varepsilon({\boldsymbol{k}})<0$ are initially occupied.
}
\label{fig_field}
\end{center}
\end{figure}

An interesting effect is the response of the HHG spectrum to a magnetic field. Here, we only consider the effect of the Zeeman term (no Lorentz force). A positive/negative $b$ shifts the up-spin band up/down and the down-spin band down/up, see Fig.~\ref{fig_field}(a,b). In our weakly filled system, a down-shift of the band means more charge in the corresponding spin channel, which translates into a higher current. Because the band shifts along the $x$ and $y$ axes are opposite, the net effect of the field on the difference spectrum $\Delta \text{HHG}=\text{HHG}(b)-\text{HHG}(-b)$ is opposite along the two axes, as is confirmed in Fig.~\ref{fig_field}(c). For a field polarized along the diagonal, $\Delta \text{HHG}$ vanishes due to symmetry. 

The sign of the difference spectrum can be understood as follows. Looking at the dispersion along $k_x$, $b>0$ pushes the spin-down band with the wide dispersion down, and the spin-up band with the narrow dispersion up (Fig.~\ref{fig_field}(a)). Since the band with the wider dispersion contributes more to the current, $\Delta \text{HHG}(b>0) > 0$ in the $x$ direction. Along $k_y$, the situation is opposite (panel (b)) and the band with the wide dispersion gets depopulated, which leads to $\Delta \text{HHG}(b>0) < 0$.  

\begin{figure}[t]
\begin{center}
\includegraphics[angle=0, width=0.32\columnwidth]{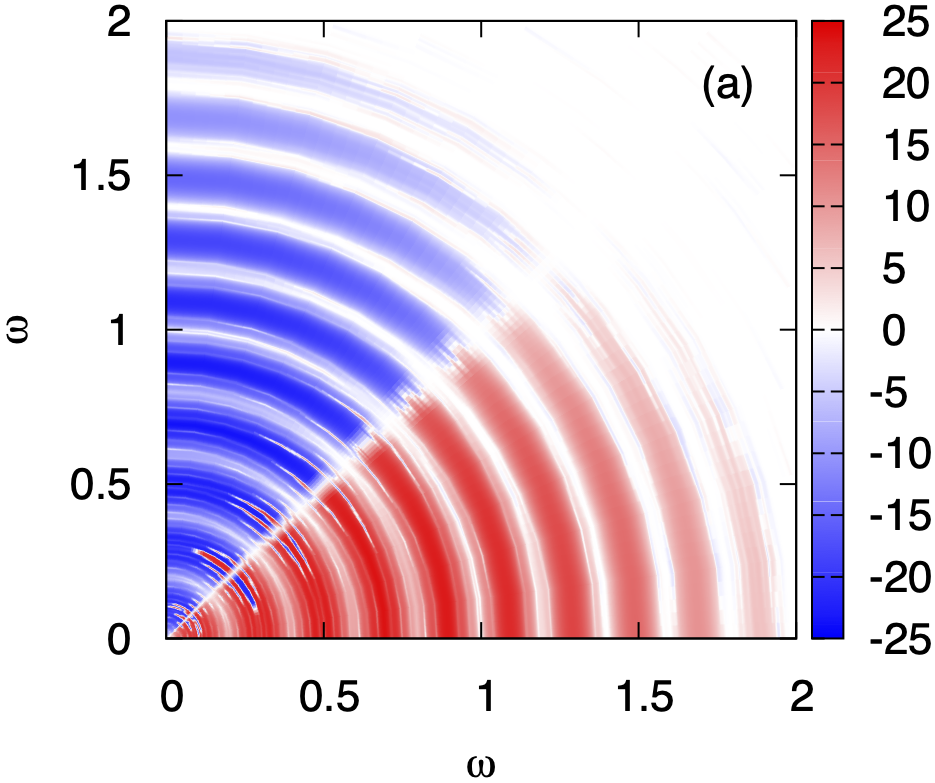} \hspace{10mm}
\includegraphics[angle=0, width=0.32\columnwidth]{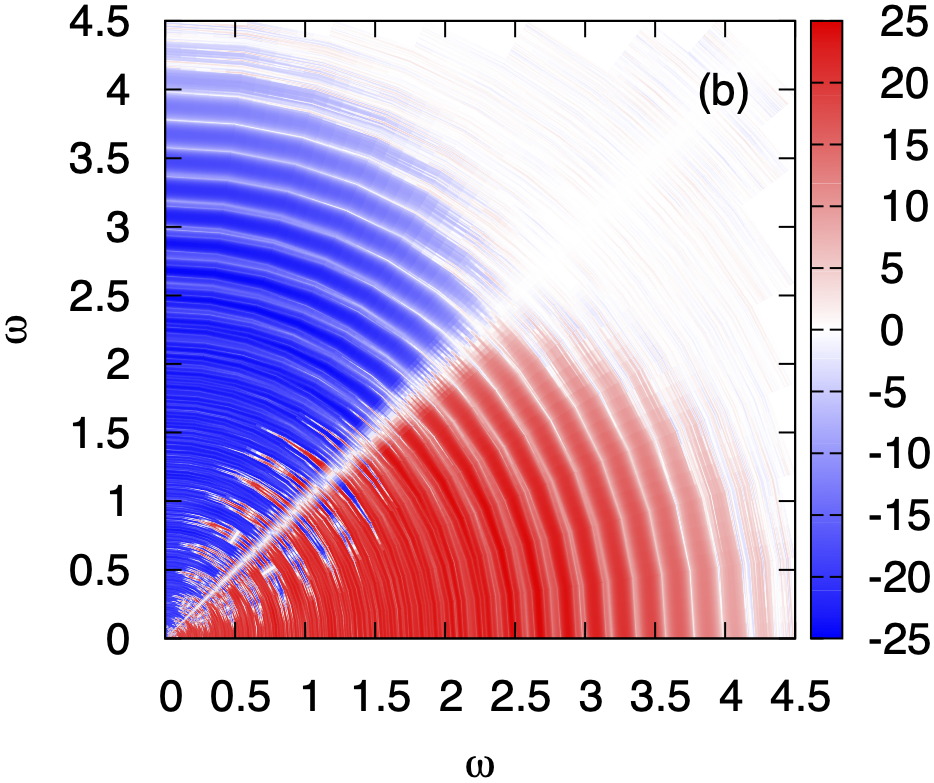}
\caption{ 
Magnetic field dependence $\Delta\text{HHG}(b)=\text{HHG}(b)-\text{HHG}(-b)$ for $b=0.01$ on a logarithmic scale for both positive an negative values ($\log(\Delta\text{HHG})+22$ for positive values and $-(\log(-\Delta\text{HHG})+22)$ for negative values). The emitted field parallel to the driving field is considered. 
Panel (a) shows the result for $E_0=1$ and panel (b) for $E_0=3$. 
}
\label{fig_hhg_polar}
\end{center}
\end{figure}

The difference spectra for the longitudinal component are shown as a function of polarization angle in Fig.~\ref{fig_hhg_polar} for two field amplitudes. Here, we used a small magnetic field $b=0.01$, for which $\omega_c\ll \Omega$, so that the Lorentz force (which is neglected in our calculations) should not have a significant effect on HHG. In order to use the log scale also on the negative axis, we plot $\log(\Delta\text{HHG})+22$ for positive values and $-(\log(-\Delta\text{HHG})+22)$ for negative values, setting the difference spectra with $|\Delta
\text{HHG}| \le e^{-22}$ to zero. The polarization dependence of the difference spectrum clearly reveals the $d$-wave nature of the altermagnet, which could be a way to experimentally detect altermagnetism.  

\subsubsection{Spin current high-harmonic spectrum}

\begin{figure}[t]
\begin{center}
\includegraphics[angle=0, width=0.32\columnwidth]{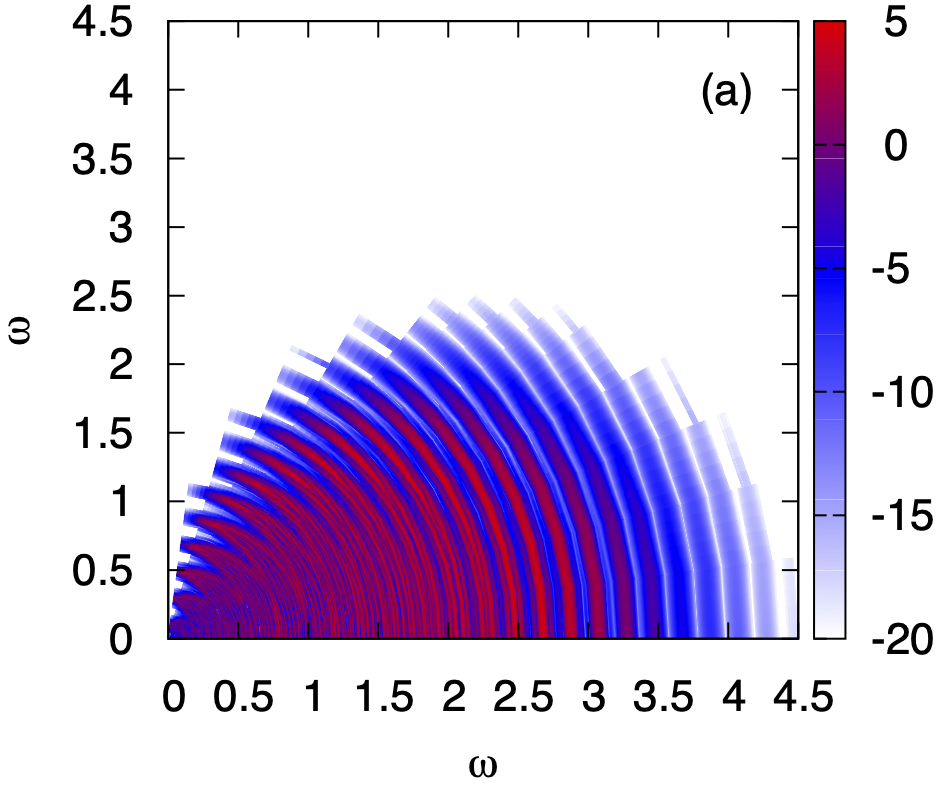}\hfill 
\includegraphics[angle=0, width=0.32\columnwidth]{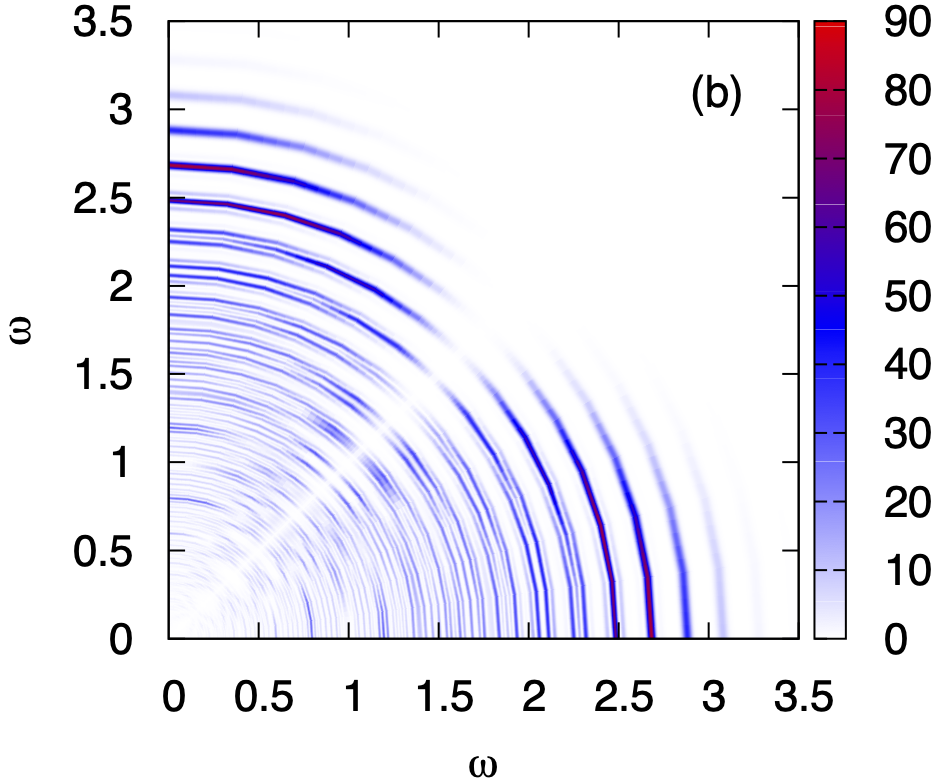}\hfill
\includegraphics[angle=0, width=0.33\columnwidth]{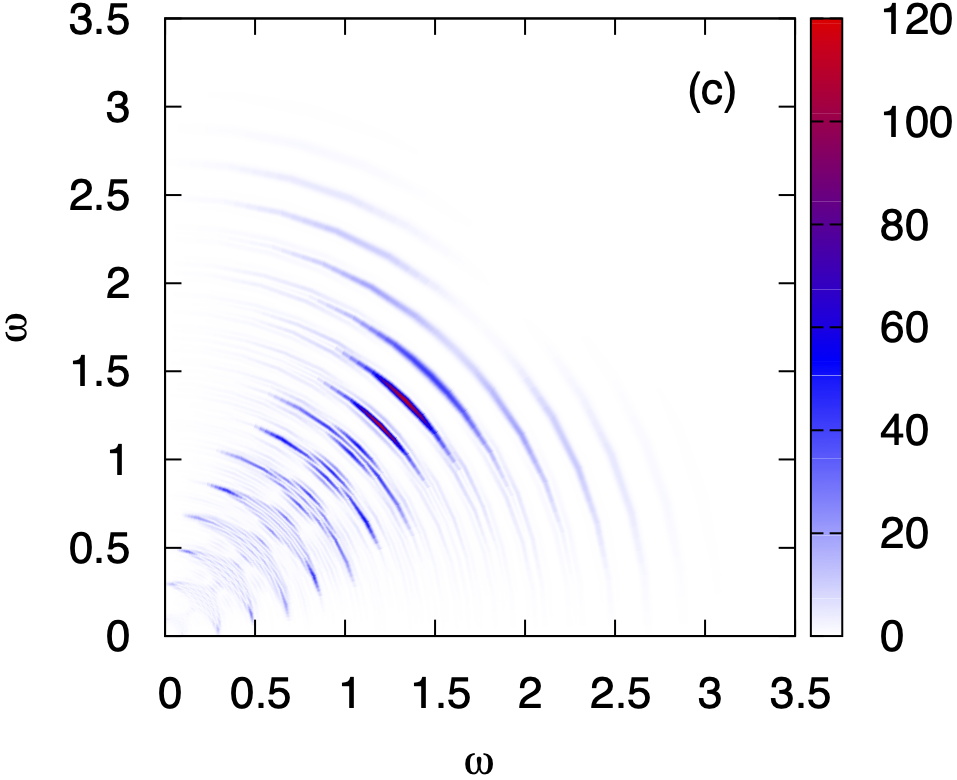}
\caption{(a) Logarithm of the spin current $J_{zx}$ high-harmonic spectrum as a function of polarization angle. 
(b) Spin current $J_{zr}$ (velocity parallel to the field) high-harmonic spectrum and (c) perpendicular spin current $J_{z\perp}$ (velocity perpendicular to the field) high-harmonic spectrum as a function of polarization angle, on a linear scale. The field strength is $E_0=3$. 
}
\label{fig_spincurrent}
\end{center}
\end{figure}

Since the spin-up and spin-down bands are not degenerate, a field polarized in a generic direction produces a spin current. Even though this current is not directly measurable, it is interesting to plot the corresponding high-harmonic spectrum, defined in a way analogous to the HHG spectrum for the charge current. The spin currents 
are calculated as in Eq.~\eqref{fig_spincurrent}. Figure~\ref{fig_spincurrent}(a) plots the logarithm of the spin high-harmonic spectrum obtained from $J_{zx}$ for different polarizations of the pump field. The result for $J_{zy}$ looks similar, but with the $x$ and $y$ axes interchanged. If the pump field has an $x$ component, and hence $v_x\ne 0$, the spectrum calculated from $J_{zx}$ becomes nonzero. Similarly the spectrum produced by $J_{zy}$ is nonzero if the field has a component along $y$. As in the case of the charge current, one can recognize the semi-circular shape of the cutoff energies, which is consistent with the projection of the electric field onto the $x$ or $y$ axes.

We define $J_{zr}=\cos(\theta)J_{zx}+\sin(\theta)J_{zy}$ as the spin current in the $\hat r$ direction (with angle $\theta$ relative to the $x$ axis). This spin current vanishes in the diagonal direction due to symmetry and thus also the $J_{zr}$-current high-harmonic spectrum vanishes along the diagonal, see Fig.~\ref{fig_spincurrent}(b) (linear scale). It reaches its highest amplitudes and cutoff energies along the $x$ and $y$ directions, where the difference in the velocities of the spin-up/down bands is maximal. Considering next the spin current perpendicular to the applied field, $J_{z\perp}=\cos(\theta+\frac{\pi}{2})J_{zx}+\sin(\theta+\frac{\pi}{2})J_{zy}$ (Fig.~\ref{fig_spincurrent}(c)), we find that its high-harmonic spectrum is maximized for the diagonal polarization, with larger amplitude but lower cutoff energy compared to $J_{zr}$. The large perpendicular spin current along the diagonal can be explained by the fact that the contributions from the two bands add up in this direction, see e. g. the illustration in Fig. 12c of Ref.~\onlinecite{Smejkal2022}.  

\subsection{Altermagnet with SOC term ($\alpha=1$)}
\label{sec:results_alpha1}

In this section we consider $h_{\boldsymbol{k}}=h^\text{alter}_{\boldsymbol{k}}+h^\text{SOC}_{\boldsymbol{k}}$ with both a nonzero spin splitting $t_a$ and spin orbit coupling $\alpha$. Unless otherwise noted, we choose $t_a=0.25$ and $\alpha=1$. Since the spin quantization axis is now locked to the $z$ axis in real/momentum space, we denote the Pauli matrices by $\sigma_{x,y,z}$. In some analyses, we furthermore add a Zeeman term with magnetic field in the $x$ direction, in which case the last term in Eq.~\eqref{eq_alter} is replaced by $b_z\sigma_z+b_x\sigma_x$. The unit of energy is again set by $t_0=1$.

\begin{figure}[t]
\begin{center}
\includegraphics[angle=-90, width=0.3\columnwidth]{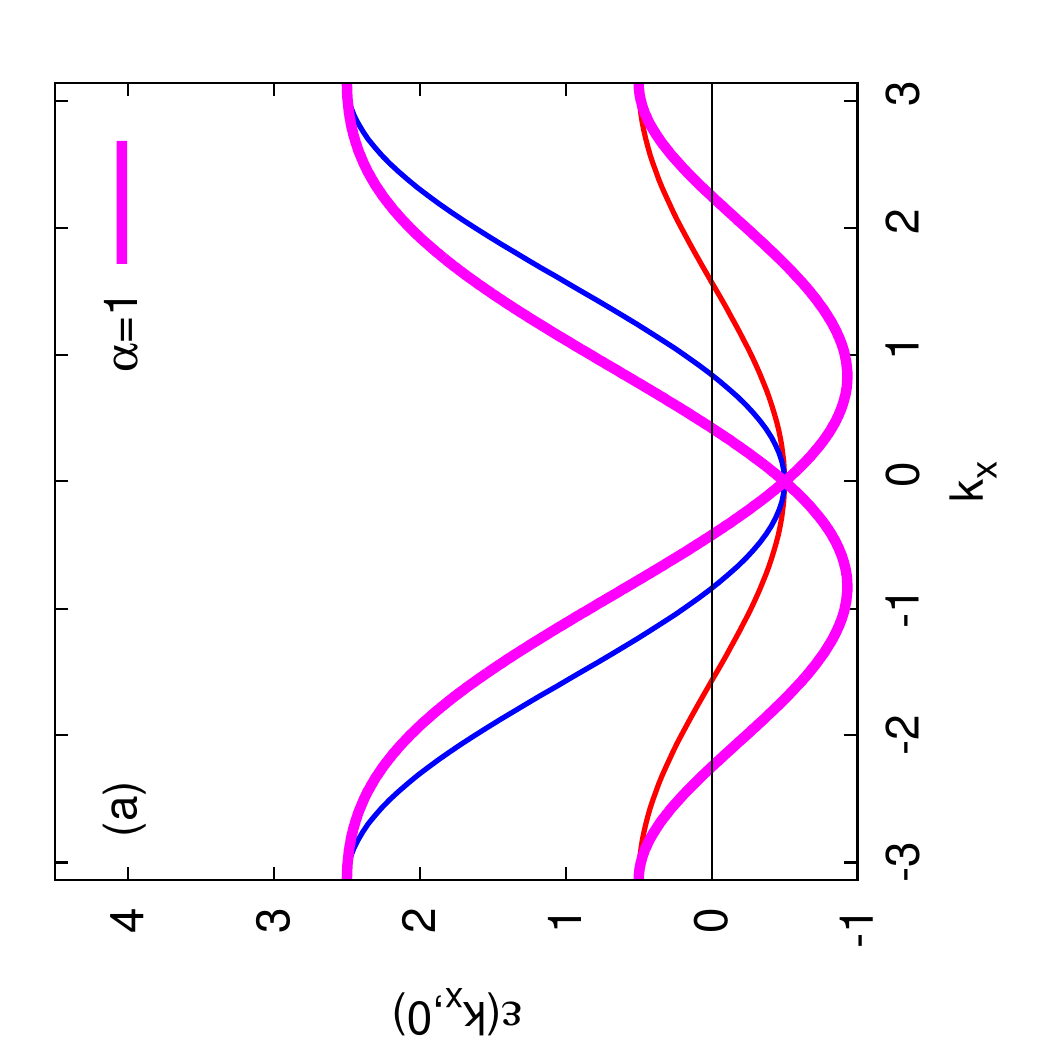}\hspace{10mm}
\includegraphics[angle=-90, width=0.3\columnwidth]{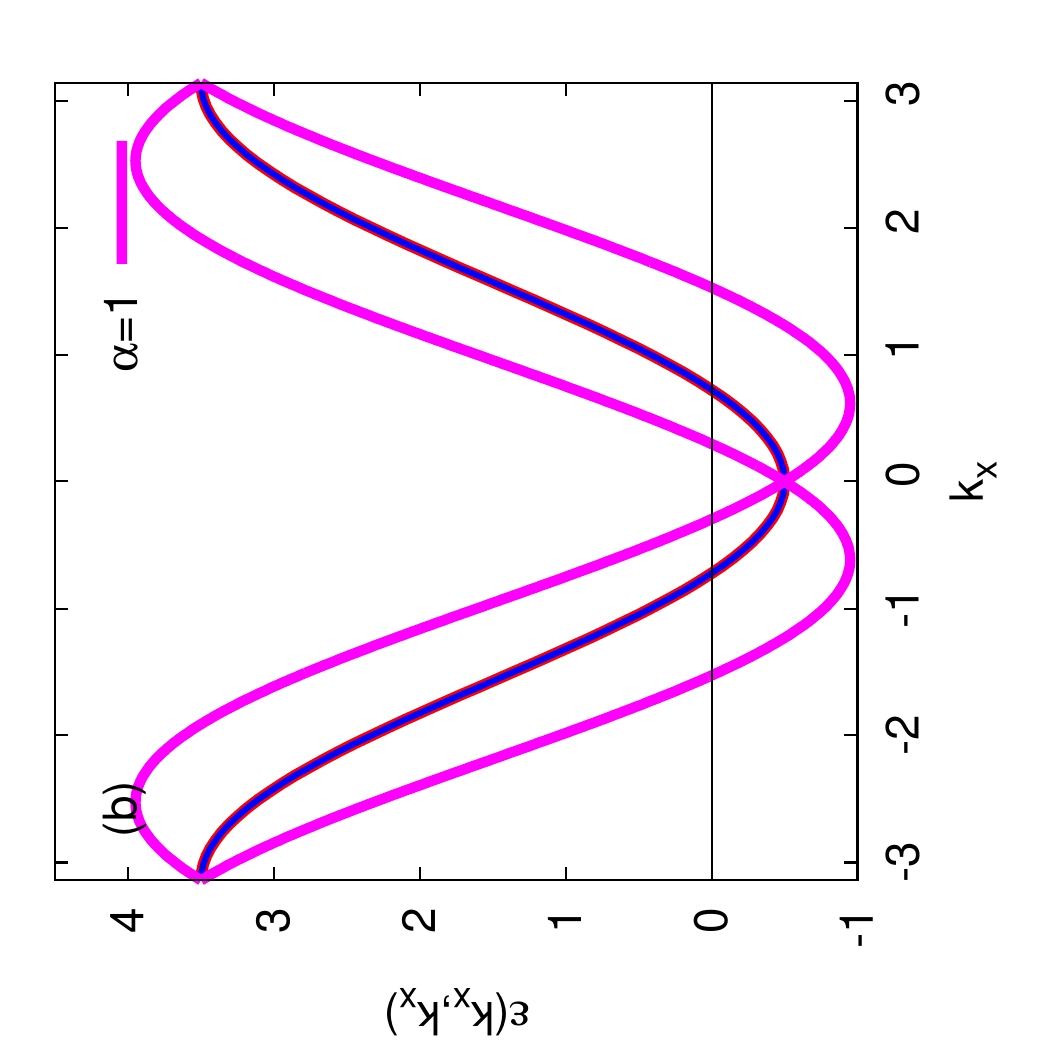}\\
\hspace{5mm}\includegraphics[angle=0, width=0.315\columnwidth]{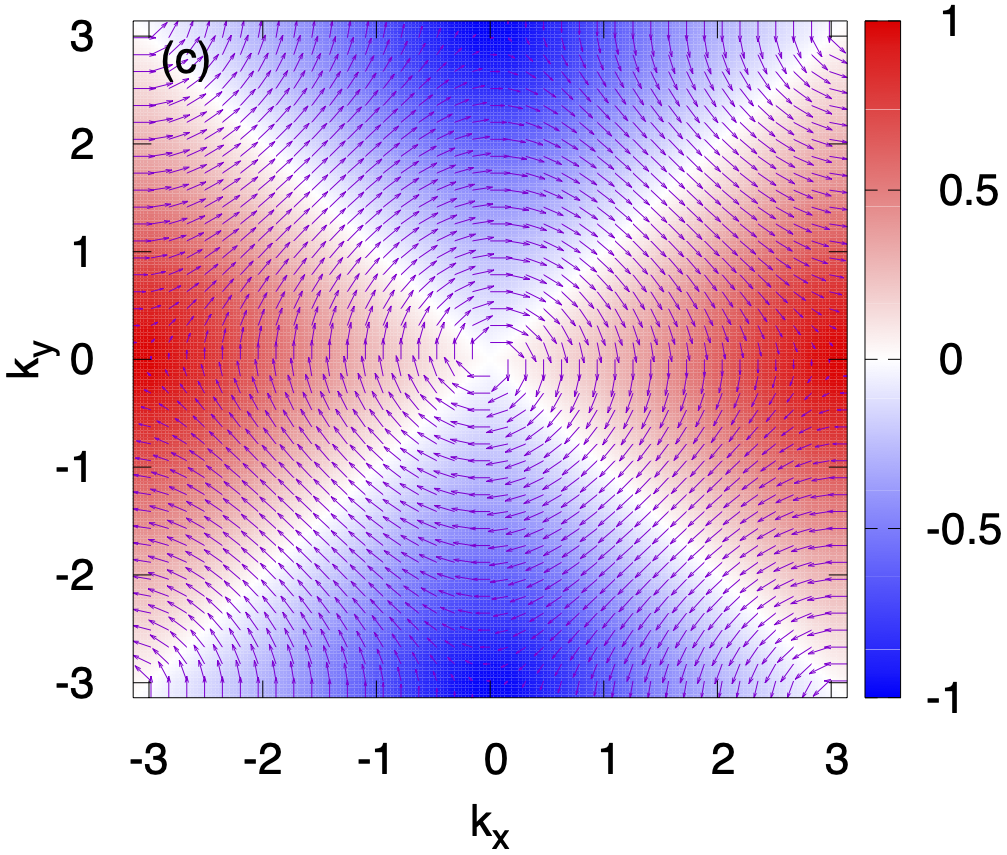}\hspace{8mm}
\includegraphics[angle=0, width=0.315\columnwidth]{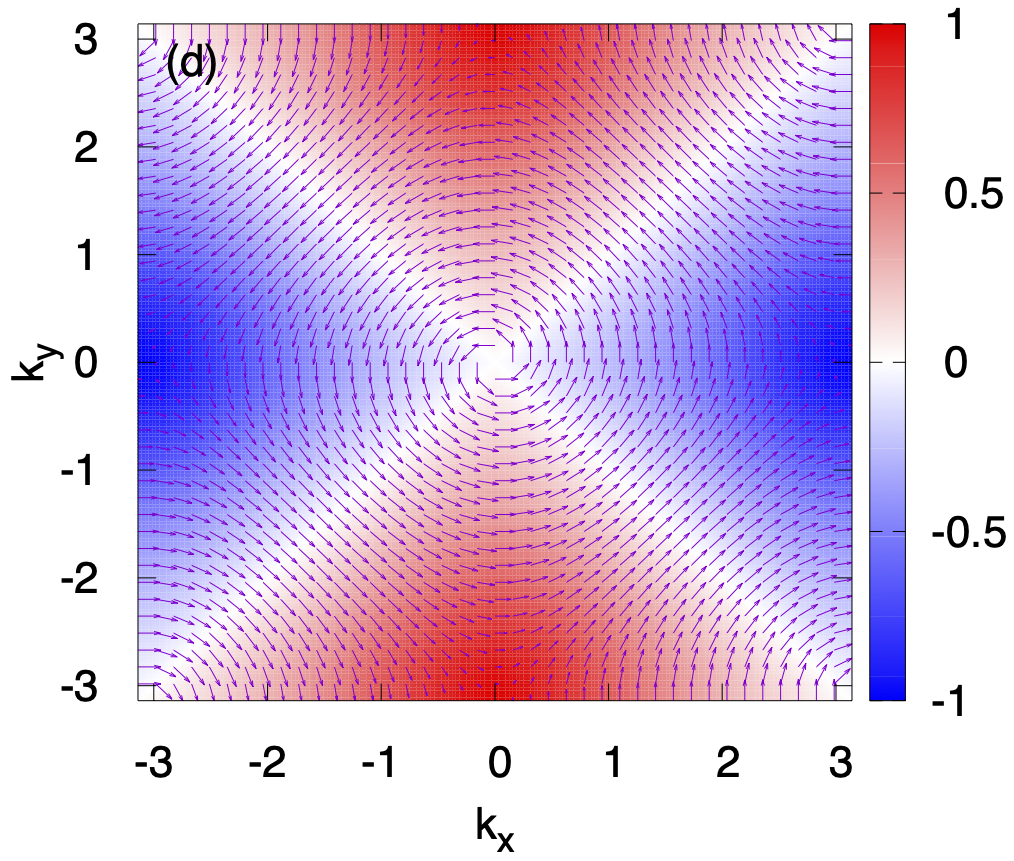}
\caption{Effect of a Rashba SOC $\alpha=1$ on the bandstructure with $t_0=1$, $t_a=0.25$, $b=0$, $\mu=0.5$. The red and blue curves in panels (a,b) show the bands for $\alpha=0$.
Panels (c,d) show the spin expectation values in the $z$ (color map) and $x,y$ (vector field) directions. 
Panel (c) is for the lower band and panel (d) for the upper band. 
}
\label{fig_soc_bands}
\end{center}
\end{figure}

The effect of the SOC term on the bandstructure is illustrated Figs.~\ref{fig_soc_bands}(a,b), where the bands for $\alpha=1$ are shown in pink, while the original spin-split bandstructure (same as in Fig.~\ref{fig_FS_bands}) is indicated by the red and blue curves. The spin textures of the lower and higher bands are shown in the lower panels. 
Here, the color gradient indicates the expectation value of $\sigma_{z}$, while the arrows indicate the expectation values of $\sigma_{x}$ and $\sigma_{y}$ (rescaled by a factor 1/5). 
Since spin is no longer a conserved quantum number, the current and HHG spectra now also contain contributions from interband processes, with a maximum energy controlled by the splitting between the bands \cite{Lysne2020}, which for a general cut in momentum space is a function of $t_a$ and $\alpha$.

\begin{figure}[t]
\begin{center}
\includegraphics[angle=0, width=0.3\columnwidth]{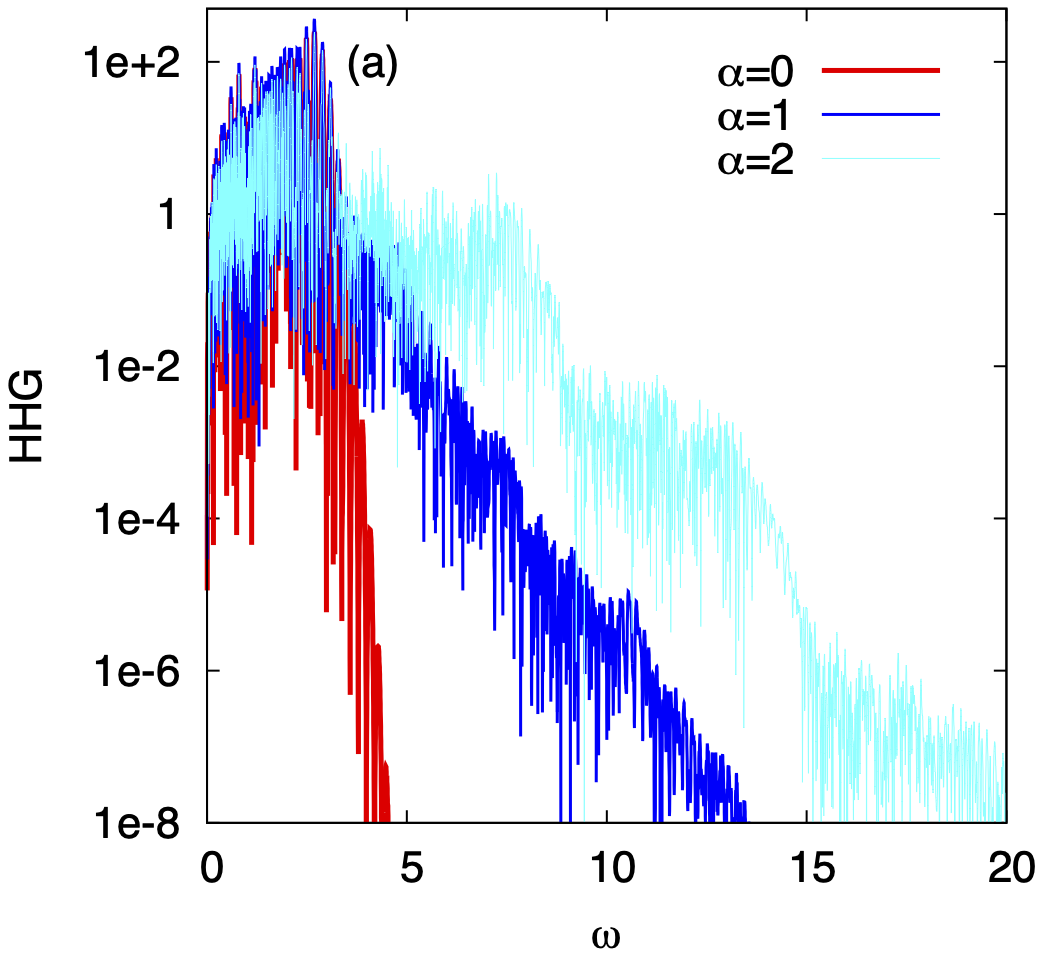}\hspace{2mm}
\includegraphics[angle=0, width=0.335\columnwidth]{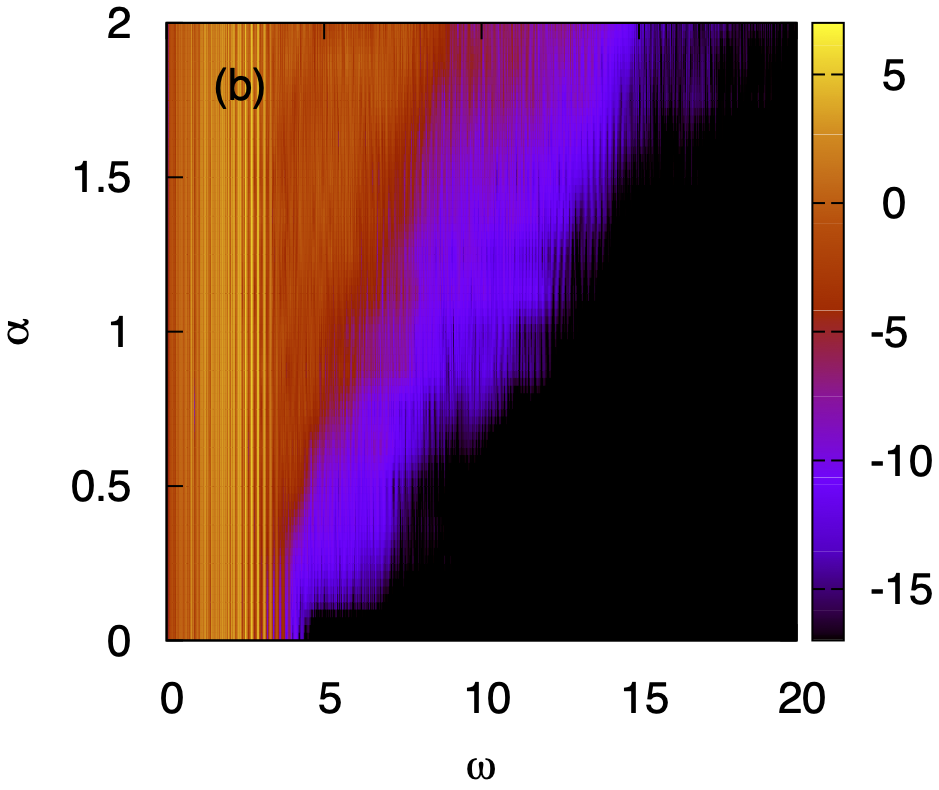}\hspace{2mm}
\includegraphics[angle=0, width=0.285\columnwidth]{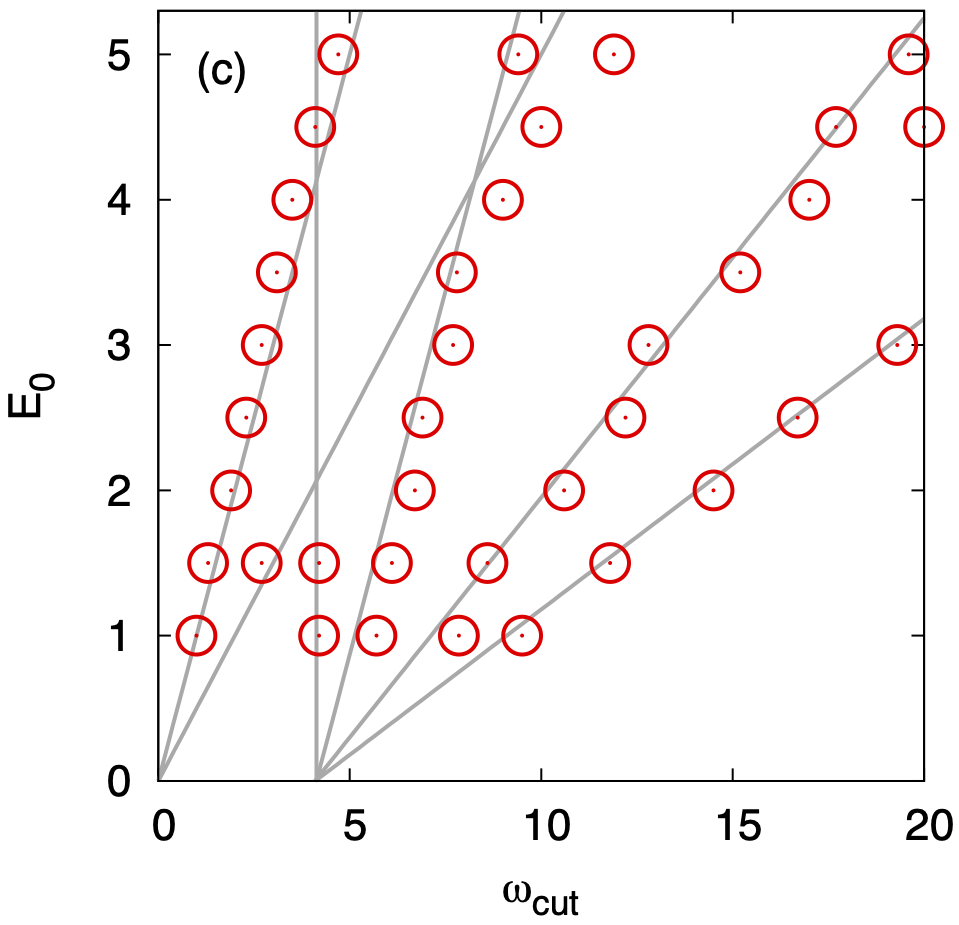}
\caption{(a) HHG spectrum for the indicated values of the SOC parameter $\alpha$ for $E_0=3$ and field polarization along $x$ ($t_a=0.25$, $b=0$, $\mu=0.5$, $500\times 500$ $k$ points). 
(b) Logarithm of the HHG spectrum as a function of $\alpha$, for the same set-up. 
(c) Cutoff energies $\omega_\text{cut}$ as a function of $E_0$ for $\alpha=2$. The two left-most gray lines are $\omega_\text{cut}=E_0$ and $2E_0$, the vertical line is $\omega_\text{cut}=4.13$ and the remaining lines are $\omega_\text{cut}=4.13+(1,3,5)E_0$. 
}
\label{fig_hhg_soc}
\end{center}
\end{figure}

\subsubsection{Longitudinal current and cutoff energies}

A selection of HHG spectra for the model with SOC term are plotted in Fig.~\ref{fig_hhg_soc} for a field with $E_0=3$ along the $x$ axis. We now find several cutoff values, a first cutoff associated with the intraband dynamics in the altermagnetic bandstructure (see red line for $\alpha=0$), and a second cutoff, increasing linearly with $\alpha$ (panel (b)), which can be explained by the additional band splitting induced by the spin orbit coupling \cite{Lysne2020}. 
Beyond the second cutoff, we find a broad tail of high-energy peaks, with indications of further plateau structures at large enough $\alpha$. 
For $\alpha=2$, panel (c) plots the all discernible cutoff energies $\omega_\text{cut}$ as a function of $E_0$ (field polarization in the $x$ direction). This figure clearly shows the previously discussed $\epsilon_\text{cut}\approx E_0$ cutoff associated with nearest-neighbor hopping, and indications of a $\epsilon_\text{cut}\approx 2E_0$ cutoff associated with next-nearest neighbor hopping. Apart from these intra-band processes, one recognizes a field-independent cutoff near $\omega_\text{cut} \approx 4.13$, which is the maximum gap for $k_y=0$. There are also further cutoffs at $\omega_\text{cut}\approx 4.13+(1,3,5)E_0$, which originate from a combination of intra-band and inter-band processes. 
Note that for the high energy cutoffs, we observe only odd multiples of $E_0$. 
Interestingly, the same property was previously found for an effective semiconductor model constructed to mimic the charge dynamics in Mott insulators~\cite{Murakami2018}.

We note that the high-energy features are sensitive to the momentum discretization. A $500\times 500$ $k$ grid was employed in these calculations and we checked that the presented results are converged (identical results are obtained for $2000\times 2000$ $k$ points). 
To suppress high-frequency oscillations after the pulse, a window function $f(t)=\cos^8(\Omega(t-t_\text{avg})/2M)$ was applied in the Fourier transform. 

\begin{figure}[t]
\begin{center}
\includegraphics[angle=0, width=0.32\columnwidth]{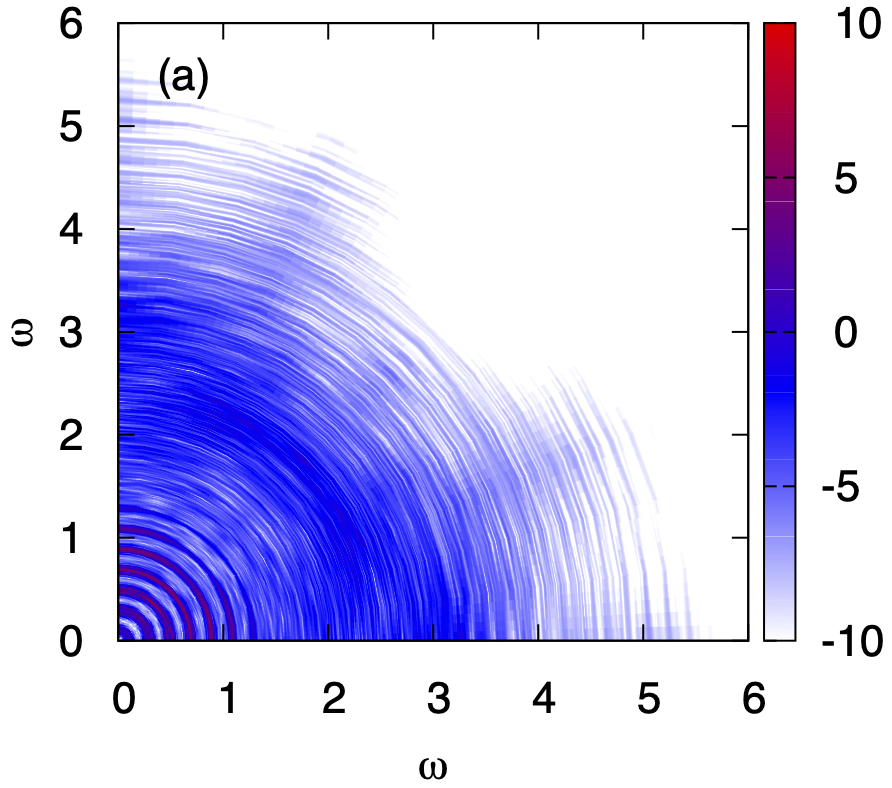}\hspace{10mm}
\includegraphics[angle=0, width=0.32\columnwidth]{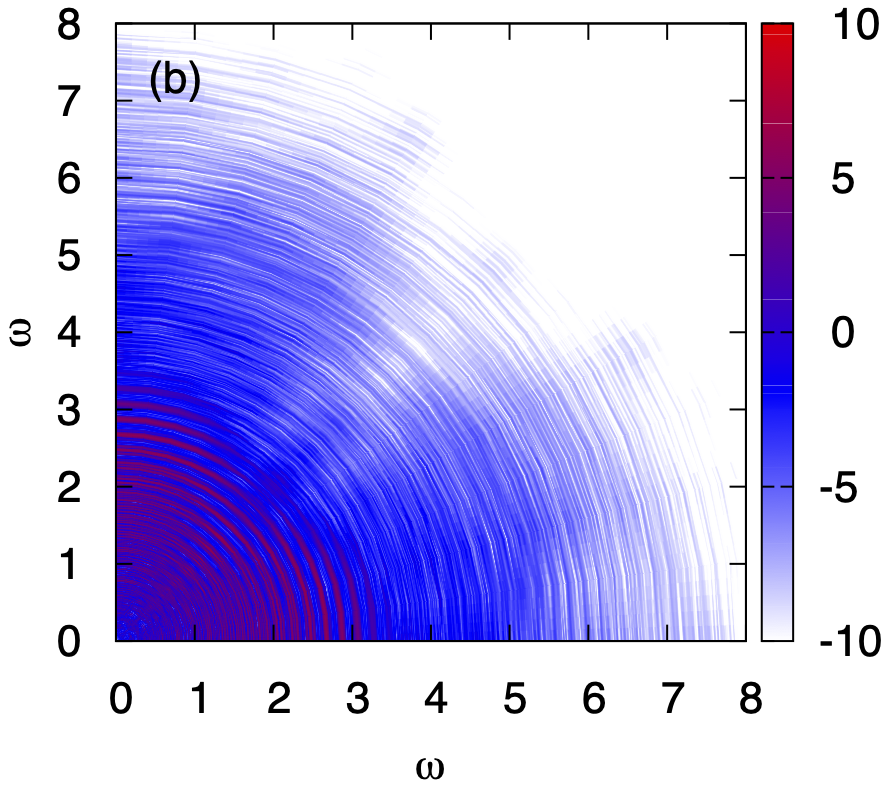}
\caption{Angular dependence of the HHG spectrum calculated from the longitudinal current for $\alpha=1$, $E_0=1$ (a) and $E_0=3$ (b). The parameters are $t_0=1$, $t_a=0.25$, $b=0$, and we use $500 \times 500$ $k$ points. 
}
\label{fig_hhg_soc_polar}
\end{center}
\end{figure}

The first two HHG plateaus are also evident in Fig.~\ref{fig_hhg_soc_polar}, which plots the spectrum as a function of the polar angle of the field. For smaller fields (panel (a)), 
the higher-energy plateau related to inter-band transitions 
has a higher intensity along the diagonal, 
where the band velocity is maximal, 
while the cutoff energies are higher along $x$ and $y$, similar to the pure altermagnet model. For higher fields (panel (b)) and in the tails of the spectrum, the highest intensity is however found away from the symmetry axes, at an angle of approximately 30 and 60 degrees. Only the edges and tails of the second plateau exhibit well defined  
harmonics at odd multiples of $\Omega$. 

\begin{figure}[t]
\begin{center}
\includegraphics[angle=0, width=0.32\columnwidth]{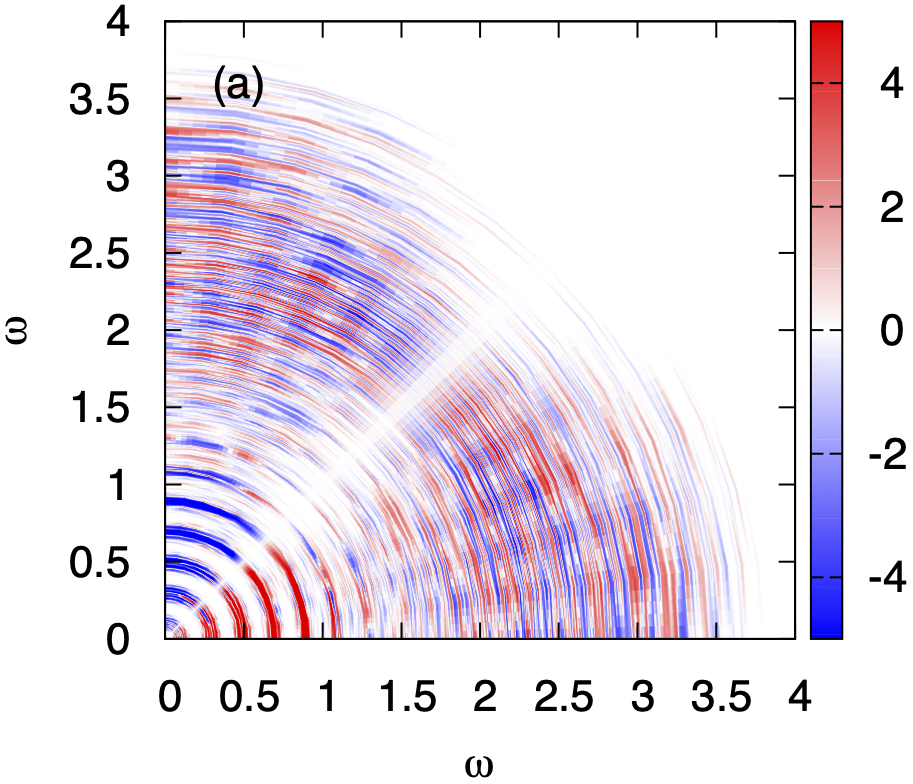}
\hspace{10mm}
\includegraphics[angle=0, width=0.32\columnwidth]{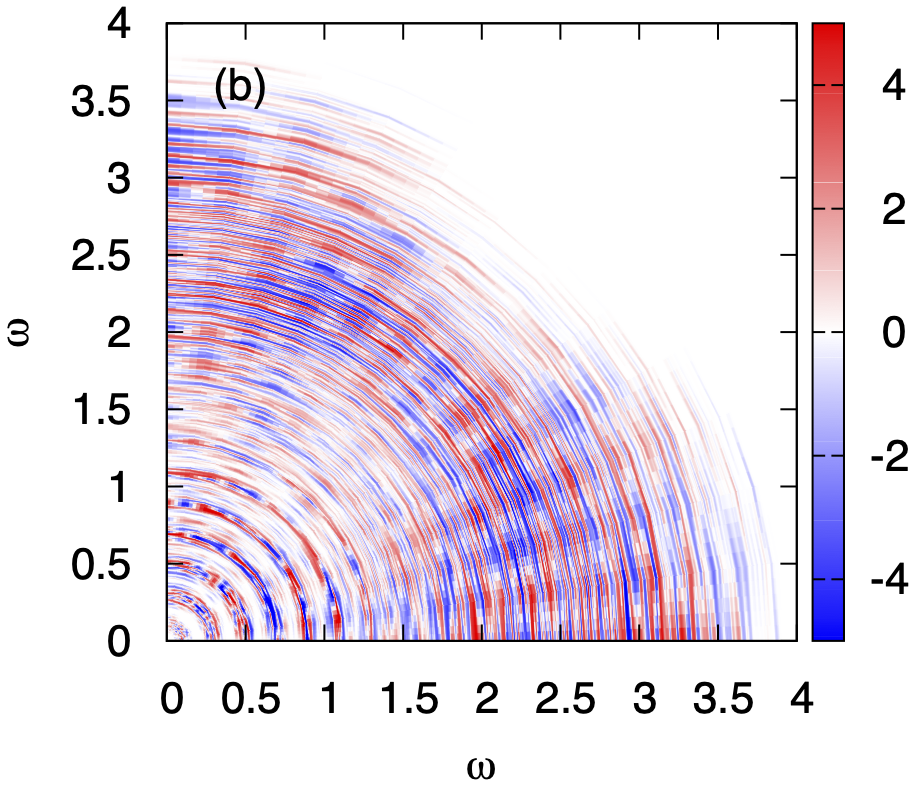}
\caption{Magnetic-field dependence of the HHG spectrum as a function of polarization angle for $E_0=1$, 
on a logarithmic scale for both positive an negative values ($\log(\Delta\text{HHG})+6$ for positive values and $-(\log(-\Delta\text{HHG})+6)$ for negative values).
(a) $\Delta_{\text{$z$,HHG}}$ for $b=\pm 0.01$ along $z$ and (b) $\Delta_{\text{$x$,HHG}}$ for $b=\pm 0.01$ along $x$. Both panels are for $t_0=1$, $t_a=0.25$, $\alpha=1$ and $500\times 500$ $k$ points. 
}
\label{fig_hhg_soc_b}
\end{center}
\end{figure}

Figure~\ref{fig_hhg_soc_b} illustrates the magnetic field dependence of the model with $\alpha=1$, for a field in the $z$ direction (panel (a)) and, for comparison, in the $x$ direction (panel (b)). In the $b_z$ case, the low-energy peaks originating from the intraband dynamics show the characteristic $d$-wave structure in $\Delta_{\text{$z$,HHG}}=\text{HHG}(b_z=0.01)-\text{HHG}(b_z=-0.01)$, with a positive/negative difference below/above the diagonal. The higher-energy signal associated with interband transitions also has a $d$-wave type sign change at the diagonal, but it features a complicated sign structure within each domain. For the field in the $x$-direction there is no simple sign pattern in the angular dependence of $\Delta_\text{$x$,HHG}=\text{HHG}(b_x=0.01)-\text{HHG}(b_x=-0.01)$.

\begin{table}
\begin{tabular}{lll}
model & transverse current \hspace{5mm} & CD in HHG \\ 
\hline
altermagnet & no & no  \\
altermagnet+$b_z$ & no & no \\
altermagnet+$b_x$ &  no & no\\
SOC & no & no \\
SOC+$b_z$ & yes & yes\\
SOC+$b_x$ & yes 
& yes\\
altermagnet+SOC & yes (nl) 
& yes \\
altermagnet+SOC+$b_z$ & yes & yes \\
altermagnet+SOC+$b_x$ \hspace{5mm} & yes 
& yes 
\end{tabular}
\caption{Transverse current (for an electric field along $x$) and existence of a CD in HHG. In the model column, ``altermagnet" refers to $t_a>0$, ``SOC" to $\alpha>0$ and ``$b_{x,z}$" to a field $b>0$ in the $x$ or $z$ direction. In the case of the altermagnet+SOC model, the transverse current scales like $E_0^3$, and hence is nonlinear (nl). The other transverse currents are linear in $E_0$ for weak enough fields. 
}
\label{tab_transverse}
\end{table}

\subsubsection{Transverse current and circular dichroism}

We now consider linearly and circularly polarized field pulses, and investigate the presence of transverse currents and circular dichroism in the current response and HHG. The linearly polarized field is in the $x$ direction, while for the circularly polarized field, we choose the components $E_x$ and $E_y$ as in Eq.~\eqref{eq_field}, with the phases $\phi_x=0$, $\phi_y=\pm\frac{\pi}{2}$.   
The results for various models are summarized 
in Tab.~\ref{tab_transverse}. For the chosen phases $\phi_i$, a circular dichroism (CD) is present if the $y$-components of the currents do not sum up to zero for field pulses with left and right circular polarization. 
If this is the case, also the HHG spectra for left and right circularly polarized pulses become nondegenerate. 

We find no transverse current in the pure altermagnet model (with or without magnetic field) and in the SOC model without magnetic field. This can be explained by the fact that these models are symmetric under the mirror operation along $y$ in real/momentum space, combined with a $\pi$ rotation around $y$ in spin space ($k_y \rightarrow -k_y$ and $\sigma_x \rightarrow -\sigma_x$, $\sigma_z \rightarrow -\sigma_z$, $\sigma_y \rightarrow \sigma_y$), while the velocity in the $y$ direction flips the sign. 

\begin{figure}[t]
\begin{center}
\includegraphics[angle=-90, width=0.32\columnwidth]{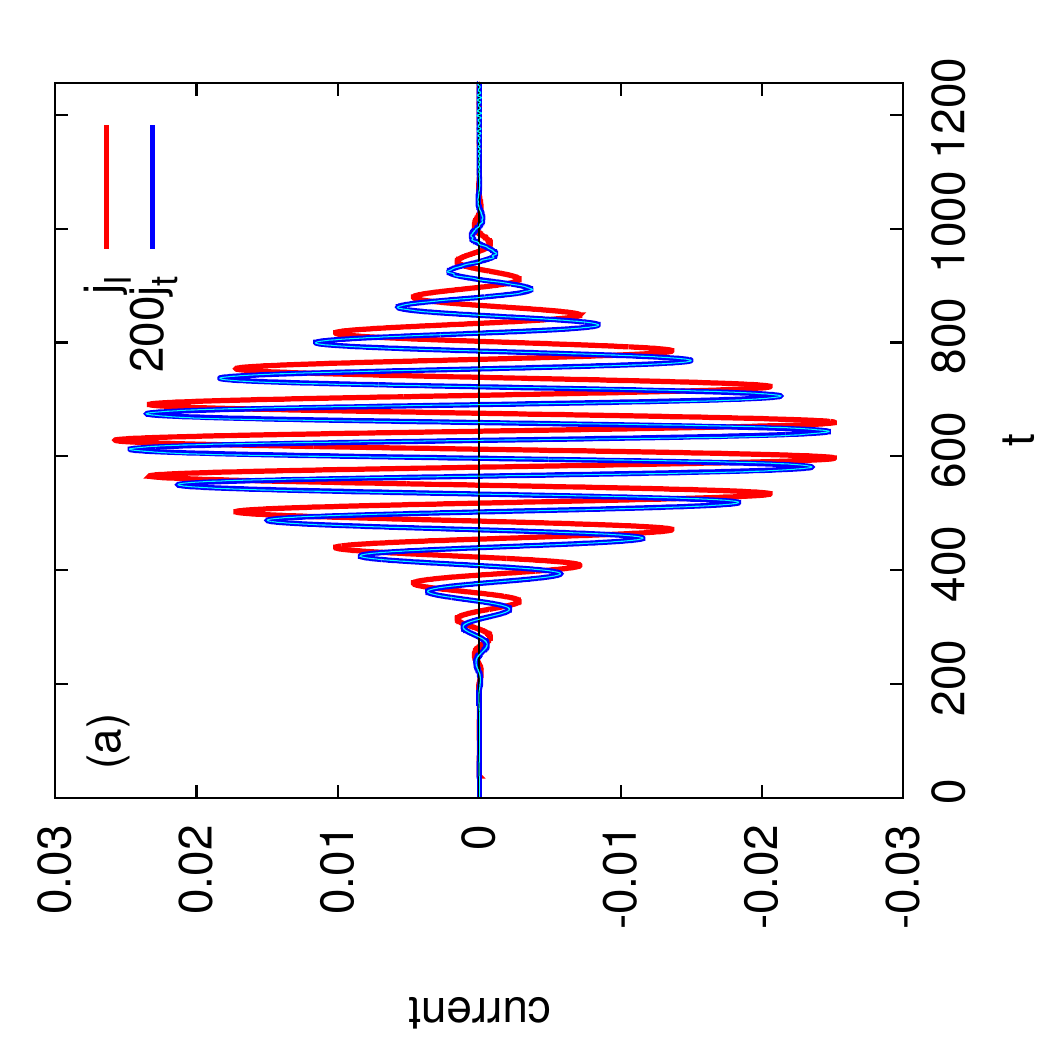}\hfill
\includegraphics[angle=-90, width=0.32\columnwidth]{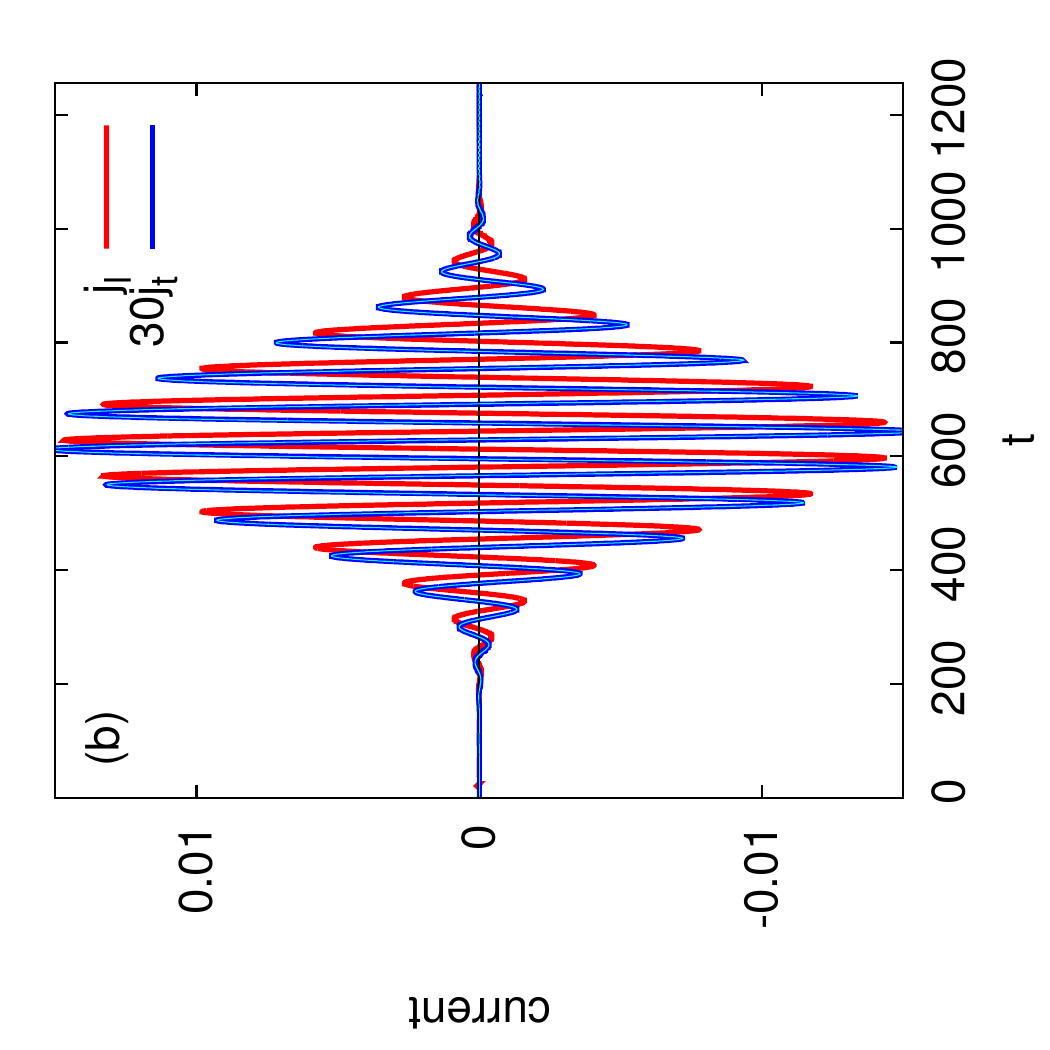}\hfill
\includegraphics[angle=-90, width=0.32\columnwidth]{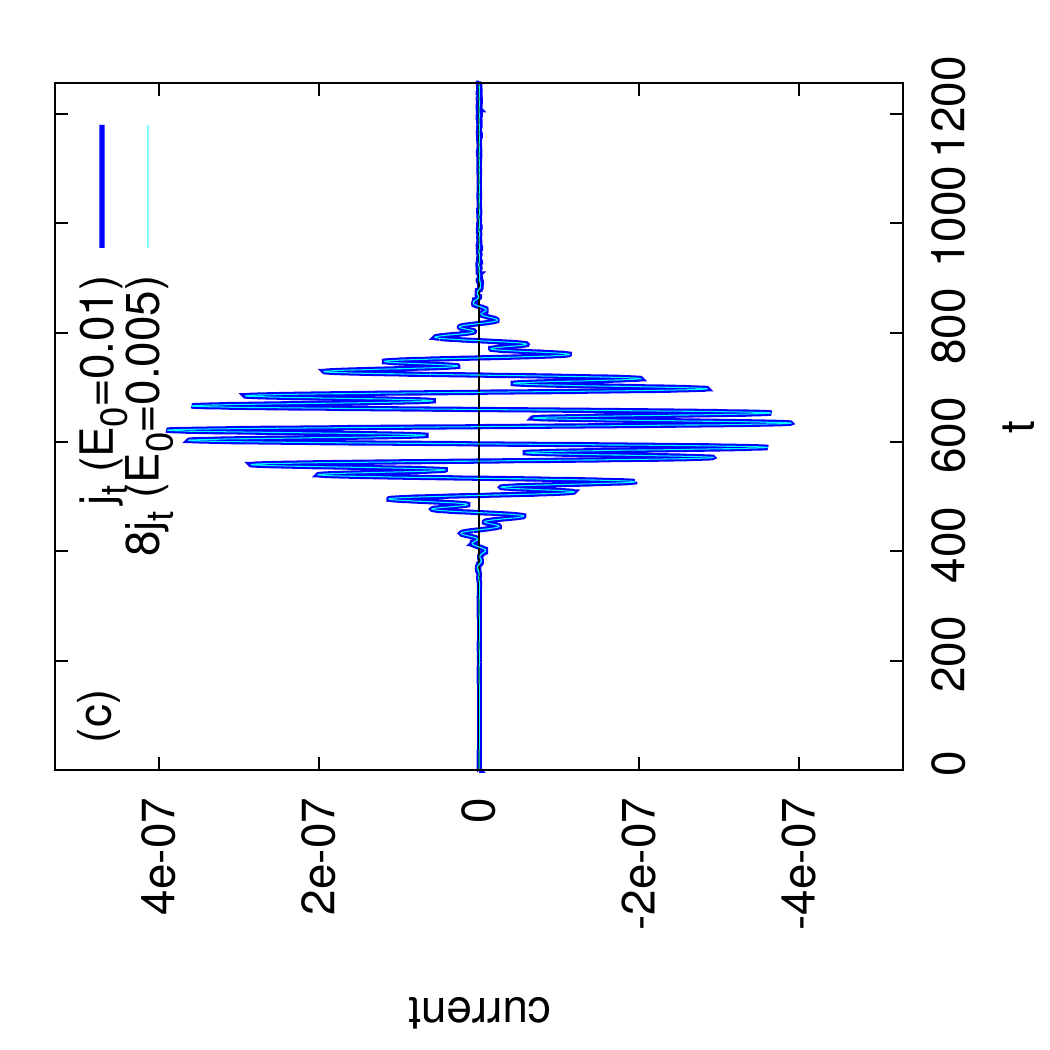}
\caption{Panels (a) and (b): Longitudinal ($j_l$) and transverse ($j_t$) currents in the altermagnet+SOC model with $\alpha=1$, $t_a=0.25$, $b_z=0.1$ (a) and $b_x=1$ (b). The pulse amplitude is $E_0=0.01$ and the polarization is in the $x$ direction. The transverse current has been rescaled by a factor 200 (a) or 30 (b) to match the scales. The light blue lines show the transverse currents for $E_0=0.02$, with half the rescaling factor, i.e. 100 (a) and 15 (b). In the topological models with nonzero $b$ field, $j_t\propto E_0$ for small enough $E_0$.
Panel (c): transverse current $j_t$ for the altermagnet+SOC model with $b=0$ and field in the $x$ direction. In this model, $j_t\propto E_0^3$ in the weak-field regime, and the current exhibits higher harmonics.  
}
\label{fig_transverse}
\end{center}
\end{figure}

As explained in Ref.~\onlinecite{Rao2024}, the SOC and altermagnet+SOC models are topological (have nonzero Chern number) in the presence of a nonzero $b_z$ field, while the altermagnet+SOC model also becomes topological in the presence of a $b_x$ field. This is consistent with our observation of a nonvanishing transverse current in these models (Tab.~\ref{tab_transverse}). 
For small enough field amplitudes, the transverse currents are proportional to the driving field $E_0$, with a prefactor determined by $b$, i.e. the effect is linear. 
This can be seen from the almost perfect match between the dark blue and light blue lines in Fig.~\ref{fig_transverse}(a,b), which correspond to $E_0=0.01$ and $E_0=0.02$, respectively, and whose scaling differs by a factor of two. The data also show that in this linear regime, the transverse current has a phase shift of $\pi/2$ relative to the longitudinal current. As shown in Tab.~\ref{tab_transverse}, the systems with nonvanishing transverse currents also exhibit a CD in the HHG spectrum. 

In contrast to the SOC and altermagnet models, the altermagnet+SOC model shows a transverse current even in the absence of a magnetic field, 
but in this case, the effect is nonlinear in the field (transverse current proportional to $E_0^3$, see Fig.~\ref{fig_transverse}(c)). 
This result implies that in a system with nonzero spin-orbit coupling, the presence of a CD in the HHG spectra can be an indication for an altermagnetic character of the underlying bandstructure. 

Let us note that for a generic field direction, the pure altermagnet model also shows nonlinear transverse charge currents proportional to $E_0^3$. However, these currents vanish for fields along the symmetry axes (e. g. polarization along $x$, as considered above), and the pure altermagnet model also does not exhibit a CD in the HHG spectrum.

\section{Conclusions}
\label{sec:conclusions}

We studied the high-harmonic response of an altermagnetic metal driven by a many-cycle electric field pulse with linear or circular polarization. In the pure altermagnet model, spin is conserved and only interband currents contribute to the HHG signal. Without magnetic field, the HHG response is similar to that of a conventional metal, except that the two spin components contribute differently to the charge current, and that the system hence also exhibits spin currents. The longitudinal charge current spectra reach their maximum intensity for fields polarized along the diagonal (where the spin up/down bands are degenerate and the dispersion is largest), while the cutoff energy is highest for fields along the $x$ and $y$ directions. In the case of the transverse spin current, the maximum intensity of the HHG signal is also found along the diagonal, where the contributions from the two bands add up. Along the $x$ and $y$ directions, the transverse spin current vanishes, while the longitudinal spin current  is maximized. 

An interesting effect specific to altermagnets is the opposite shift of the spin-up/down bands in the presence of a Zeeman field. This leads to a characteristic $d$-wave pattern in the difference spectra $\Delta\text{HHG}(b)=\text{HHG}(b)-\text{HHG}(-b)$. Measurements of $\Delta\text{HHG}(b)$ for different polarization directions could thus reveal an altermagnetic bandstructure.  

In the presence of spin-orbit coupling, interband transitions are activated, and the HHG spectra exhibit two prominent cutoff scales, a low-energy cutoff associated with intra-band dynamics (as in the pure altermagnet) and a cutoff related to the spin-orbit enhanced bandgap. In the presence of a $b_z$ field, the harmonics up to the first cutoff show the characteristic altermagnet sign structure in  $\Delta\text{HHG}(b_z)$, while the higher harmonics related to interband transitions exhibit a complicated sign pattern (although still with a $d$-wave symmetry). In the presence of a $b_x$ field our model does not exhibit any $d$-wave structure in $\Delta\text{HHG}(b_x)$. 

The models with spin-orbit coupling and magnetic fields produce a nonvanishing transverse current, even for fields along $x$, and a circular dichroism in the HHG spectrum, consistent with the nonzero Chern numbers in these models. Interestingly, even without any magnetic fields, the altermagnetic system with spin-orbit coupling shows a nonvanishing nonlinear transverse response, and a circular dichroism in HHG, even though the pure altermagnet and pure SOC models do not. In systems with non-negligible spin-orbit coupling, the observation of a circular dichroism in the HHG spectra could thus be an indication for altermagnetism. 

In this study, we considered noninteracting electrons without impurity scattering or other damping mechanisms, and we assumed that the cyclotron motion is negligible. As a follow-up project, it would be interesting to study the effects of interactions and disorder, and the effect of the Lorentz force in models with magnetic fields.

\acknowledgements

The calculations were carried out on the beo05 cluster at the University of Fribourg. Y.~M. acknowledges support by Grant-in-Aid for Scientific Research from JSPS, KAKENHI Grant Nos. JP21H05017, JP24H00191, JST CREST Grant No. JP-MJCR1901, and the RIKEN TRIP initiative RIKEN Quantum.

\end{document}